\documentclass[preprint,12pt]{elsarticle}

\usepackage{graphicx,subfigure}
\usepackage{amssymb}
\usepackage{amsmath}
\usepackage{bm}
\usepackage{lineno} 
\usepackage{natbib}
\usepackage[squaren]{SIunits}
\usepackage{multirow}
\usepackage{xcolor}

\begin{document}

\begin{frontmatter}

\title{On numerical errors to the fields surrounding a relativistically moving particle in PIC codes}
 
\author[UCLAEE,SLAC]{Xinlu Xu}
\ead{xuxinlu@slac.stanford.edu; Present address: SLAC National Accelerator Laboratory, Menlo Park, California, 94025, USA}
\author[UCLAEE]{Fei Li} 
\author[UCLAPH]{Frank S. Tsung}
\author[UCLAPH]{Thamine N. Dalichaouch} 
\author[BNU]{Weiming An}
\author[UCLAEE]{Han Wen}
\author[UCLAPH]{Viktor K. Decyk}
\author[IST,ISCTE]{Ricardo A. Fonseca}
\author[SLAC]{Mark J. Hogan}
\author[UCLAEE,UCLAPH]{Warren B. Mori}

\address[UCLAEE]{Department of Electrical Engineering, University of California Los Angeles, Los Angeles, CA 90095, USA}
\address[SLAC]{SLAC National Accelerator Laboratory, Menlo Park, California 94025, USA}
\address[UCLAPH]{Department of Physics and Astronomy, University of California Los Angeles, Los Angeles, CA 90095, USA}
\address[BNU]{Department of Astronomy, Beijing Normal University, Beijing 100875, China}
\address[IST]{GOLP/Instituto de Plasma e Fus\~ao Nuclear, Instituto Superior T\'ecnico, Universidade de Lisboa, Lisbon, Portugal}
\address[ISCTE]{ISCTE - Instituto Universit\'ario de Lisboa, 1649--026, Lisbon, Portugal}

\begin{abstract}
The particle-in-cell (PIC) method is widely used to model the self-consistent interaction between discrete particles and electromagnetic fields. It has been successfully applied to problems across plasma physics including plasma based acceleration, inertial confinement fusion, magnetically confined fusion, space physics, astrophysics, high energy density plasmas. In many cases the physics involves how relativistic particles (those with high relativistic $\gamma$ factors) are generated and interact with plasmas. However, when relativistic particles stream across the grid both in vacuum and in plasma there are many numerical issues that may arise which can lead to incorrect physics. We present a detailed analysis of how discretized Maxwell solvers used in PIC codes can lead to numerical errors to the fields that surround particles that move at relativistic speeds across the grid. Expressions for the axial electric field as integrals in $\bm{k}$ space are presented.  Two types of errors to these expressions are identified. The first arises from errors to the numerator of the integrand and leads to unphysical fields that are antisymmetric about the particle. The second arises from errors to the denominator of the integrand and lead to Cerenkov like radiation in ``vacuum". These fields are not anti-symmetric, extend behind the particle, and cause the particle to accelerate or decelerate depending on the solver and parameters. The unphysical fields are studied in detail for two representative solvers - the Yee solver and the FFT based solver. Although the Cerenkov fields are absent, the space charge fields are still present in the fundamental Brillouin zone for the FFT based solvers. In addition, the Cerenkov fields are present in higher order zones for the FFT based solvers. Comparison between the analytical solutions and OSIRIS results are presented. A solution for eliminating these unphysical fields by modifying the $\bm{k}$ operator in the axial direction is also presented. Using a customized finite difference solver, this solution was successfully implemented into OSIRIS. Results from the customized solver are also presented. This solution will be useful for a beam of particles that all move in one direction with a small angular divergence.

\end{abstract}


\end{frontmatter}


\section{Introduction}
\label{sect:intro}
The particle-in-cell (PIC) method has been well developed and widely used to model the interactions between charged particles and electromagnetic fields for over half a century \cite{dawson1983particle, hockney1988computer, birdsall1991plasma}. In this method space is broken up into discrete grids or finite size cells. The positions and velocities of finite size particles with a shape function $S(\bm{x}-\bm{x_p}(t) ) $ which can have continuous values for $\bm{x_p}(t)$ are used to deposit the currents (and/or charges) of the particles onto the corners of the grids. These are used as source terms in a discretized version of Maxwell's equations to advance the fields. The fields are used to interpolate forces onto the particles when are then advanced to new positions and momentum using the relativistic version of Newton's equations of motion.  The PIC method greatly reduces the computational cost of electrostatic problems, e.g., in electrostatic PIC codes the computational cost of $N$ particles is $O(N\ln N)$ as compared to $O(N^2)$ when using action at distance method \cite{hockney1988computer}; and makes studying many body electromagnetic problems feasible. However, the need to deposit information from the particles that have continuous positions onto discrete grid locations leads to issues with aliasing, and the use of a discretized version of Maxwell's equations can lead to errors in the dispersion relation of light even in vacuum. These difficulties lead to many well known numerical issues in the PIC method  \cite{birdsall1991plasma}. 


The PIC method has been and continues to be used  to model a variety of problems in plasma and beam physics \cite{dawson1983particle}. In some problems, the entire plasma, e.g, relativistic shocks \cite{bykov2012particle}, or a group  of plasma/beam particles have speeds where relativistic mass corrections become important, e.g., plasma based acceleration \cite{joshi2003plasma} and fast ignition \cite{craxton2015direct, betti2016inertial}. It is well known that for some Maxwell solvers the phase velocity of light in ``vacuum" is less than the speed of light so that relativistically moving particles can radiate unphysical Cerenkov radiation \cite{godfrey2013numerical}. The use of the grid and finite difference time operators essentially means that the grid can be viewed as a medium where the dispersion relation of light is modified. In addition, there has been recent work on identifying and mitigating or eliminating what is referred to as the numerical Cerenkov instability (NCI) \cite{xu2013numerical, yu2014modeling, godfrey2014numerical, godfrey2014suppressing, yu2015elimination, godfrey2015improved, yu2015mitigation, yu2016enabling, kirchen2016stable, lehe2016elimination, li2017controlling} that arises from the coupling between the electromagnetic and plasma beam modes. These schemes include using a variety of different Maxwell solvers including solvers that customize the representation of spatial derivatives in wave number space \cite{li2017controlling}.  The NCI cannot be eliminated by simply having the phase velocity of light be equal to or greater than the speed of light because of aliasing.  

In this article, we address another issue for studying relativistic particles using a PIC code. The fields surrounding an electron (point of finite size) moving with constant speed are well known. They can be obtained by calculating the fields in vacuum and then Lorentz transforming them into the moving frame. The axial electric field is a Lorentz invariant while the fields perpendicular (transverse) to the direction of motion are increased by $\gamma$. Thus, the axial electric field is relatively small for highly relativistic electrons. We call these space charge like fields and they are antisymmetric about the particle, so they do not lead to self-forces. We show that for the PIC algorithm numerical errors lead to space charge fields that are orders of magnitude larger than the correct values.  In addition, depending on the choice of the solver, the particle will radiate thereby creating both axial and transverse fields around the particle. We call these Cerenkov like fields and they are not antisymmetric (they can extend behind the particle), so they can create `self-forces' on the particle. We will show that depending on the solver these forces can be either accelerating or decelerating.  

These numerical errors to the fields that surround a single particle lead to distortions to the evolution of a beam of particles. This issue can be problematic when modeling how a relativistic particle beam propagates in vacuum and in a plasma. Specifically, we have found that when the current profile rises rapidly these fields can lead to unphysical energy spread and modulations to a beam. In fact, it was through an investigation into the cause of this unphysical distortion of relativistic beams that we have identified the unphysical space charge and Cerenkov fields that surround a particle. 

This article is organized as follows. In section \ref{sect:theoretical}, we begin with a formal derivation of the fields generated by a finite size particle moving with constant speed across a grid. The fields are first calculated in Fourier space. A general expression for the axial electric field is given as an integral in wave number space. This expression includes effects associated with aliasing as a sum over all Brillouin zones. Two sources of numerical errors are identified in a fraction found in the integrand. The first is from the numerator and the second is from the denominator. This expression is analyzed for the Yee and FFT (spectral) based solvers. For the Yee solver, zeros in the denominator exist in the fundamental Brillouin zone and these can lead to Cerenkov radiation (and thus "self-forces"). These self-forces slow the particle down. Errors in the numerator arise for high $\gamma$ leading to significant errors to the space charge fields. For the first Brillouin zone there is no Cerenkov radiation but there are still spurious space charge fields. 

We next show that for an FFT based solver there is no Cerenkov radiation in the fundamental Brillouin zone but there are still spurious space charge fields. However, there are Cerenkov like fields (zeros in the denominator) in the first Brillouin zones and thus self-forces, as well as spurious space charge fields. The self-forces accelerate the particles.  Comparison between the analytical fields and those obtained from OSIRIS are given and there is good agreement. Our analyses are done for a single particle which can be extended to a bunch of particles through convolution with the distribution function of the particles. 

In section \ref{sect:solution}, we then propose a solution that can significantly reduce the errors to the fields that surround the particle. The proposed solution is a modification to the $\bm{k}$-space operator of derivatives along the axial direction.
Essentially the $\bm{k}$-space operator in $\hat{x}_1$ is replaced with $\frac{\mathrm{sin} (k_1 dt/2)}{dt/2}$. Such a solver has perfect dispersion for light moving along $\hat{x}_1$ and this is achieved by modifying the differential operator in real space to match the time operator. This is essential for removing numerical errors to the space charge fields. Although perfect dispersion in vacuum is also achieved in the PSATD method \cite{haber1973advances, birdsall1991plasma, vay2013domain} for waves moving in all directions, it is not as effective at eliminating the space charge forces as the proposed method. This is because for the PSATD solver perfect dispersion is achieved by effectively modifying the time domain operator. For the proposed solution there are small errors to the space charge fields from the first Brillouin zones. The proposed solution can be easily implemented into a FFT based solver and we show how it can be implemented into a customized finite difference solver using an over specified higher order solver whose coefficients are chosen to minimize errors from the desired $\bm{k}$-space operators. We then present results obtained from OSIRIS using the proposed customized solver. These results are in close agreement with the analytical results. Results from OSIRIS simulations of a drifting electron beam show a dramatic difference between using a standard vs. customized finite difference solver. A summary is given in section \ref{sect:summary}. Finally, three appendices are included. In \ref{sec: AppendixA} details for the form of the charge density of a free streaming particle including aliasing is given; in \ref{sec: psatd} it is shown that the the fields surrounding a particle from the PSATD algorithm will be similar to those for the FFT solver; in \ref{sec: cus} details of the proposed customized solver are given; and finally in \ref{sec:integrations} details about the complex integrations are given.

\section{Theoretical Analysis}
\label{sect:theoretical}
\subsection{General expressions of the EM fields induced by free-streaming particles}
\label{subsect:general}
In this paper we are concerned with the fields that surround a charged particle free streaming along the $x_1$ direction on a grid. We assume that the simulation grid is infinitely long (which is equivalent to a large box with open boundary conditions). The simulation time is also assumed to be infinitely long and that the fields reach ``steady state". We show later that in fact the fields can oscillate over time as the particle moves due to aliasing on the grid. The corresponding Fourier transform of these discrete non-periodic physical quantities defined on the grids are continuous and periodic in the $\omega-\bm{k}$ space. 

We start from the discretized form of Maxwell's equations, i.e., Faraday's and Ampere's Law, which are used in the PIC method to advance the fields (Gauss's law is satisfied by ensuring charge conservation),
\begin{align}
\mathrm{d}_t \bm{B} &= - \bm{\mathrm{d}}_E \times \bm{E} \nonumber \\
\mathrm{d}_t \bm{E} &= \bm{\mathrm{d}}_B \times \bm{B}- \bm{J}
\end{align}
which upon Fourier transforming gives,
\begin{align}
[\omega]_t \bm{\tilde{B}} &=  [\bm{k}]_E \times \bm{\tilde{E}}	\label{eq: Faraday}\\
[\omega]_t \bm{\tilde{E}} &= - [\bm{k}]_B \times  \bm{\tilde{B}}  -  i \bm{\tilde{J}} \label{eq: Ampere}
\end{align}
where $[\bm{k}]_E$ and $[\bm{k}]_B$ are the $\bm{k}$-space operators for the choice of the discretized form for the spatial derivatives used on the $\bm{E}$ and $\bm{B}$ fields in Maxwell equations. We allow for different forms of the operators be used in Faraday's and Ampere's Laws. Here we use $[.]$ exclusively to indicate the discrete operator as in  previous work \cite{xu2013numerical, yu2015elimination}, e.g., for the leap frog operator in the particle push and a second order finite difference operator, $[\omega]\equiv \frac{sin (\omega dt/2)}{dt/2}$ and $[k]_x \equiv \frac{sin (k_x dx/2)}{dx/2}$. The details can be found in \ref{sec: cus}. Applying $[\bm{k}]_B \times$ to both side of Eq. (\ref{eq: Faraday}) and using Eq. (\ref{eq: Ampere}), the coupled wave equation for $\bm{\tilde{E}}$ is obtained as
\begin{align}
\left( [\omega]_t^2 - [\bm{k}]_E \cdot [\bm{k}]_B + [\bm{k}]_E[\bm{k}]_B  \cdot \right) \bm{\tilde{E}} =  - i [\omega]_t \bm{\tilde{J}} \label{eq:wave}
\end{align}

The current deposition scheme is complicated in the PIC codes and will need to be corrected to ensure charge conversation in different ways corresponding to the choice of the differential operators in the solvers, thus we choose to write the expressions for the fields in terms of the charge density and not the current. 
We substitute Gauss's law,

\begin{align}
i [\bm{k}]_B \cdot \bm{\tilde{E}} = \tilde{\rho} \label{eq:Gauss}
\end{align}
into Eq. (\ref{eq:wave}) and use the continuity equation to rewrite the current  for a particle moving only in $\hat x_1$ in terms of $\rho$,  $\bm{\tilde J}=\hat x_1 \frac{[\omega]_t}{[k]_{B1}} \rho$. This provides an expression for $\bm{\tilde{ E}}$ in terms only of $\rho$ whose components  are, 
\begin{align}
 \tilde{E}_1 &=   - \frac{i}{[k]_{B1}}\frac{  [\omega]_t^2 - [k]_{E1}[k]_{B1} } {   [\omega]_t^2 - [\bm{k}]_E [\bm{k}]_B } \tilde{\rho}, \tilde{E}_{2} = i \frac{ [k]_{E2} } {  [\omega]_t^2 - [\bm{k}]_E [\bm{k}]_B }\tilde{\rho}, \tilde{E}_3 =  i \frac{ [k]_{E3} } {  [\omega]_t^2 - [\bm{k}]_E [\bm{k}]_B }\tilde{\rho}. \nonumber \\
 \end{align}
 Expressions for the components of $\bm{\tilde {B}}$ can be obtained by using these expressions for $\bm{\tilde {E}}$ in Faraday's law,
\begin{align}
\tilde{B}_1 &= 0, \tilde{B}_2 = - i \frac{ [k]_{E3}}{[k]_{B1}}\frac{[\omega]_t }{  [\omega]_t^2 -  [\bm{k}]_E [\bm{k}]_B }\tilde{\rho}, \tilde{B}_3 =   i \frac{ [k]_{E2}}{[k]_{B1}}\frac{[\omega]_t }{  [\omega]_t^2 -  [\bm{k}]_E [\bm{k}]_B }\tilde{\rho}
\label{eq: EM}
\end{align}

The charge density of the free streaming particles at time step  $n$ can be expressed as $\rho^n(x_1,x_2,x_3)=\rho^0(x_1 - \beta n dt, x_2,x_3)$ and an expression for the Fourier transform of the charge density, $\tilde{\rho}$, on the grid points is derived in \ref{sec: AppendixA}. 

For the remainder of this paper, we concentrate on the axial component of the electric field $E_1$ as this is the component that can do work on the particle. If we substitute $\tilde{\rho}$ from Eq. (\ref{eq:rho}) into Eq. (\ref{eq: EM}), we obtain,
\begin{align}
\tilde{E}_1 = - \frac{i }{ [k]_{B1} }\frac{  [\omega]_t^2 - [k]_{E1}[k]_{B1} } {\left( [\omega]_t^2 -  [\bm{k}]_E [\bm{k}]_B \right) } \frac{2\pi}{dt dx_1 dx_2 dx_3}\sum_{\mu,\bm{\nu}} S(\bm{k}')  \tilde{\rho}^0 (\bm{k}') \delta\left( \omega + \mu \omega_g - \beta k'_{1}  \right) \label{E1_kspace}
\end{align}
where $\tilde{\rho}^0(\bm{k})$ is the Fourier transform in space of the initial charge distribution of the particles, $S(\bm{k})$ is the shape function of the particles in the $\bm{k}$ space, $\omega_g=\frac{2\pi}{dt}$ and $k'_{1,2,3} = k_{1,2,3} +\nu_{1,2,3} k_{g1,2,3}$ where $k_{g1,2,3}=\frac{2\pi}{d x_{1,2,3}}$. Inverting the Fourier transform of $\tilde{E}_1 (\omega, \bm{k})$ back to time and  space (discrete values of time and space) leads to,
\begin{align}
{E}^n_{1, i_1 , i_2, i_3 }  &= -\frac{1}{(2\pi)^3}  \int_{-\bm{k_g}/2}^{\bm{k_g}/2} \mathrm{d}\bm{k} \int_{-\omega_g/2}^{\omega_g/2} \mathrm{d}\omega \frac{i }{ [k]_{B1} }\frac{  [\omega]_t^2 - [k]_{E1}[k]_{B1} } {\left( [\omega]_t^2 -  [\bm{k}]_E [\bm{k}]_B \right) } \frac{2\pi}{dt dx_1 dx_2 dx_3}\sum_{\mu,\bm{\nu}} S(\bm{k}')  \tilde{\rho}^0 (\bm{k}') \nonumber \\
&\delta\left( \omega + \mu \omega_g - \beta k'_{1}  \right) \mathrm{exp}\left( i k_1 i_1 dx_1 + i k_2 i_2 dx_2 + i k_3 i_3 dx_3 - i \omega n dt \right) \nonumber \\
&=   -\frac{1}{(2\pi)^3}  \int_{-\bm{k_g}/2}^{\bm{k_g}/2} \mathrm{d}\bm{k}  \frac{ i}{ [k]_{B1}} \sum_{\bm{\nu}}\frac{ [\beta k'_1 ]^2_t - [k]_{E1}[k]_{B1} } {  [\beta k'_1 ]^2_t  - [\bm{k}]_E [\bm{k}]_B }  S(\bm{k'})   \tilde{\rho}^0 (\bm{k}') \nonumber \\
 &\mathrm{exp}\left[ i k_1( i_1 dx_1  - \beta n dt)+ i k_2 i_2 dx_2 + i k_3 i_3 dx_3 \right] \mathrm{exp}( - i\beta \nu_1 k_{g1} n dt)  \label{eq:E1}
\end{align}
where the summation over $\mu$ is removed because for each $\nu_1$ there is only one $\mu$ which satisfies $-\omega_g/2 < \beta \left( k_{1} +\nu_1 k_{g1} \right) - \mu \omega_g \leq \omega_g/2$ for $k_1$ in the fundamental Bruillouin region. Note that the phase terms in the exponential functions will have additional terms like $\pm \frac{1}{2}i k_1 dx_1$, $\pm \frac{1}{2}i k_2 dx_2$, or $\pm \frac{1}{2}i k_3 dx_3$ \cite{xu2013numerical} when the staggered grids are used. 

In the continuous limit, it is straightforward to show that Eq. (\ref{eq:E1}) reduces to the well known result for a moving charge $q$ \cite{Jackson},
\begin{align}
{E}_{1 }  (t, \bm{x}) &=   -\frac{q}{(2\pi)^3}  \int_{-\infty}^{+\infty} \mathrm{d}\bm{k} \frac{ i(1-\beta^2) k_1} {  (1-\beta^2)k_1^2 +  k_2^2 + k_3^2}  \mathrm{exp}\left[ i k_1(x_1  - \beta t) + i k_2 x_2 + i k_3 x_3\right] \nonumber \\
&= \frac{q}{4\pi} \frac{\gamma (x_1-\beta t)}{\left[ \gamma^2(x_1-\beta t)^2 + x_2^2 + x_3^2 \right]^{{3}/{2}}}
\end{align}

It can be  seen  that  in the continuous limit the numerator in the integrand has a factor $1-\beta^2=1/\gamma^2$ that is very small for relativistic particles. It is important that the expression for the PIC algorithm also scale this way. For comparison to the PIC results to be presented later we also give the continuous result for two dimensions,
\begin{align}
{E}_{1 }  (t, x_1, x_2) &=   -\frac{\lambda}{(2\pi)^2}  \int_{-\infty}^{+\infty} \mathrm{d}\bm{k} \frac{ i(1-\beta^2) k_1} {  (1-\beta^2)k_1^2 +  k_2^2 }  \mathrm{exp}\left[ i k_1(x_1  - \beta t) + i k_2 x_2\right] \nonumber \\
&= 2\lambda \frac{\gamma (x_1-\beta t)}{ \gamma^2(x_1-\beta t)^2 + x_2^2 }
\end{align}
where $\lambda$ is the charge per unit length in the translationally invariant direction.
When using a grid, it can be easily shown that  the difference between $[.]_t^2$ and $[.]_{E1}[.]_{B1}$ in Eq. (\ref{eq:E1}) can typically dominate  $1/\gamma^2$ in the numerator of the integrand for most of the frequency range.

As an example, consider the Yee solver for which the numerator factor normalized to $k_1^2$ is 
\begin{align}
\frac{ [\beta k_1]_t^2 - [k]_1^2 }{k_1^2} \approx -\frac{1}{\gamma^2} + \left[  \frac{dx_1^2-dt^2}{12}  + \frac{dt^2}{6\gamma^2}  + O\left(\frac{1}{\gamma^4}\right) \right] k_1^2  +  O(k_1^4 )\label{eq: taylor expansion}.
\end{align} 
It can easily be seen that only for  $|k_1\frac{dx_1}{2}| \ll \left( {\gamma}\sqrt{\frac{1-dt^2/dx_1^2}{3} + \frac{2}{3\gamma^2}  \frac{dt^2}{dx_1^2} } \right)^{-1} \sim \frac{1}{\gamma}$  will the difference between $\beta^2$ and 1 dominate. Thus the fields surrounding the particle will be purely numerical in most of the $k_1$ frequency region for relativistic particles. The wave number components contained in the particles distribution which are not resolved by the grids will contribute to the fields through the aliasing. 

The phase factor $\mathrm{exp}(-i \beta \nu_1 k_{g1} n dt)$ in Eq. (\ref{eq:E1}) leads to variation of the fields with the time step $n$ as the particle moves between grid points. Here, we ignore this term as it is a common factor to the field expression. This common factor can vary as the particle moves between grid points.
To further simplify the analysis, a point charge which initially resides at the origin is considered, i.e., $\tilde{\rho}^0(\bm{k}') = q$. Note in our description (see \ref{sec: AppendixA}), $\rho$ represents the charge density of particle centers and $S(\bm{x})$ represents the shape of each particle if it was centered at $\bm{x}=0$. Therefore, the variables $\nu_{2,3}$ only appear in the shape function and the summation over $\nu_2, \nu_3$ depends on the particle shapes. If linear shapes are assumed in the transverse directions, i.e., $S=\mathrm{sin}^2 (k)/k^2$,  then  $\sum_{\nu_2,\nu_3} S(k_1+\nu_1 k_{g1} , k_2 + \nu_2 k_{g2}, k_3 + \nu_2 k_{g3}) = S_1 (k_1+\nu_1 k_{g1})$ where $\sum_{\nu=-\infty}^{+\infty} \frac{\mathrm{sin}^2(k+\nu \pi)}{ (k+\nu \pi)^2} = 1$ is used and $S_1$ is the shape function along the $x_1$ direction. 

Clearly the fraction in the integrand reduces to unity in the 1D limit. Therefore, the numerical effects addressed in this paper only exist in multi-dimensions. For simplicity we only consider the 2D case. In addition, in the continuous limit  the $E_1$ field vanishes as $\beta$ approaches unity.  We therefore consider the limit of  $\beta=1$ because in this limit the resulting fields are all due to numerical errors. We carry out the integral in $\bm{k}$ space in 2D Cartesian geometry for $\beta=1$ to examine in detail the numerical errors for the $E_1$ field for a relativistic speed on the grids, 
\begin{align}
{E}^n_{1, i_1 , i_2} &=  - \frac{q}{(2\pi)^2} \int_{- \frac{k_{g1}}{2}}^{\frac{k_{g1}}{2}} \int_{-\frac{k_{g2}}{2}}^{\frac{k_{g2}}{2}}\mathrm{d}k_1 \mathrm{d}k_2  \frac{ i }{ [k]_{B1} }\sum_{\nu_1} S_1(k'_{1})\frac{ [ k'_1 ]^2_t - [k]_{E1}[k]_{B1} } {  [ k'_1 ]^2_t  - [k]_{E1}[k]_{B1} - [k]_{E2}[k]_{B2} }   \nonumber \\
 &\mathrm{exp}\left( i k_1  i'_1 dx_1 + ik_2 i_2 dx_2 \right)
 \label{eq: E12D}
\end{align}
where $ i'_1=i_1 -N$ is the grid number relative to the point charge and $N\equiv \beta n \frac{\mathrm{d} t}{\mathrm{d}x_1}$ is an integer. In PIC codes the shape function is chosen so that it rapidly approaches zero as $|k_1|$ approaches and then exceeds $k_{g1}$. Therefore, contributions from each Brillouin zone are progressively smaller. It what follows, we only consider the contributions from the fundamental Brillouin zone $\nu_1=0$ and the first aliasing zones $\nu_1=\pm1$.

When performing the integrals,  the poles of the denominator of the integrand, i.e., the zeros of the function $[k_1 + \nu_1 k_{g1}]_t^2 - [k]_{E1}[k]_{B1} - [k]_{E2}[k]_{B2}$,  modify the character of the fields. This is analogous to the continuous limit where poles of the denominator lead to Cerenkov radiation in a medium where the phase velocity of light is less than $c$. The value of the denominator depends on the grid sizes, the time step, the solver type (the forms of $[.]_{t,1,2,3}$) and the value of $\nu_1$ (fundamental or aliasing zones).
Generally, there are three different scenarios depending on the values of the two key parameters $r\equiv dx_2/dx_1$ and $\kappa\equiv dx_1/dt$. The first scenario is that for all $k_1$ in the fundamental zone ($|k_1|\leq k_{g1}/2$) the denominator can vanish for some $k_2$, i.e., the integration function has singularities when integrating over $k_2$. In this case, the fields will have a wake structure analogous to Cerenkov radiation. In the second scenario,  for all $k_1$ in the fundamental zone, there is no $k_2$ for which the denominator vanishes. In this case the fields around the particle are antisymmetric and keep up with the particle. We call these space charge like (they are like the fields in the continuous limit) as compared to Cerenkov like. In the third scenario the fields are all mixed between space charge and Cerenkov like. For some ranges of $k_1$ the denominator can be zero while for other ranges of $k_1$ the denominator cannot vanish. We note that the physical condition which leads to Cerenkov radiation, $\beta_{ph}<1$, does not work exactly for numerical grids, where $\beta_{ph} \equiv \omega / k$ is the phase velocity of the EM waves.

\subsection{The fields with the Yee solver}
The Yee solver is currently the most common choice in PIC codes owing to it being fast, stable, relatively accurate and easy to be parallelized. For the Yee solver, the  frequency and wave number operators are
\begin{align}
[\omega]_t = \frac{\mathrm{sin}(\omega \frac{dt}{2})}{ \left(\frac{dt}{2}\right) }, [k]_{E1} = [k]_{B1} = \frac{\mathrm{sin}(k_1 \frac{dx_1}{2})}{\left(\frac{dx_1}{2}\right)}, [k]_{E2} =[k]_{B2} = \frac{\mathrm{sin}(k_2 \frac{dx_2}{2})}{\left(\frac{dx_2}{2}\right)} 
\end{align}

The contribution from the fundamental Brillouin zone to the $E_1$ field is
\begin{align}
{E}^n_{1, i_1 , i_2 } (\nu_1=0)  &=  - \frac{q}{(2\pi)^2} \int_{- \frac{k_{g1}}{2}}^{\frac{k_{g1}}{2}} \int_{-\frac{k_{g2}}{2}}^{\frac{k_{g2}}{2}} \mathrm{d}k_1 \mathrm{d}k_2  \frac{ i S(k_{1})}{[k]_1 } \frac{ [ k_1]^2_t - [k]_1^2 } {  [ k_1 ]^2_t  - [k]_1^2 - [k]_2^2 } \nonumber \\
&\mathrm{exp}\left[ i k_1 dx_1 \left(i'_1  + \frac{1}{2} \right) + i k_2 i_2 dx_2 \right] 
\end{align}
where the staggered grids are used. 

\begin{figure}[htbp]
\begin{center}
\includegraphics[width=0.5\textwidth]{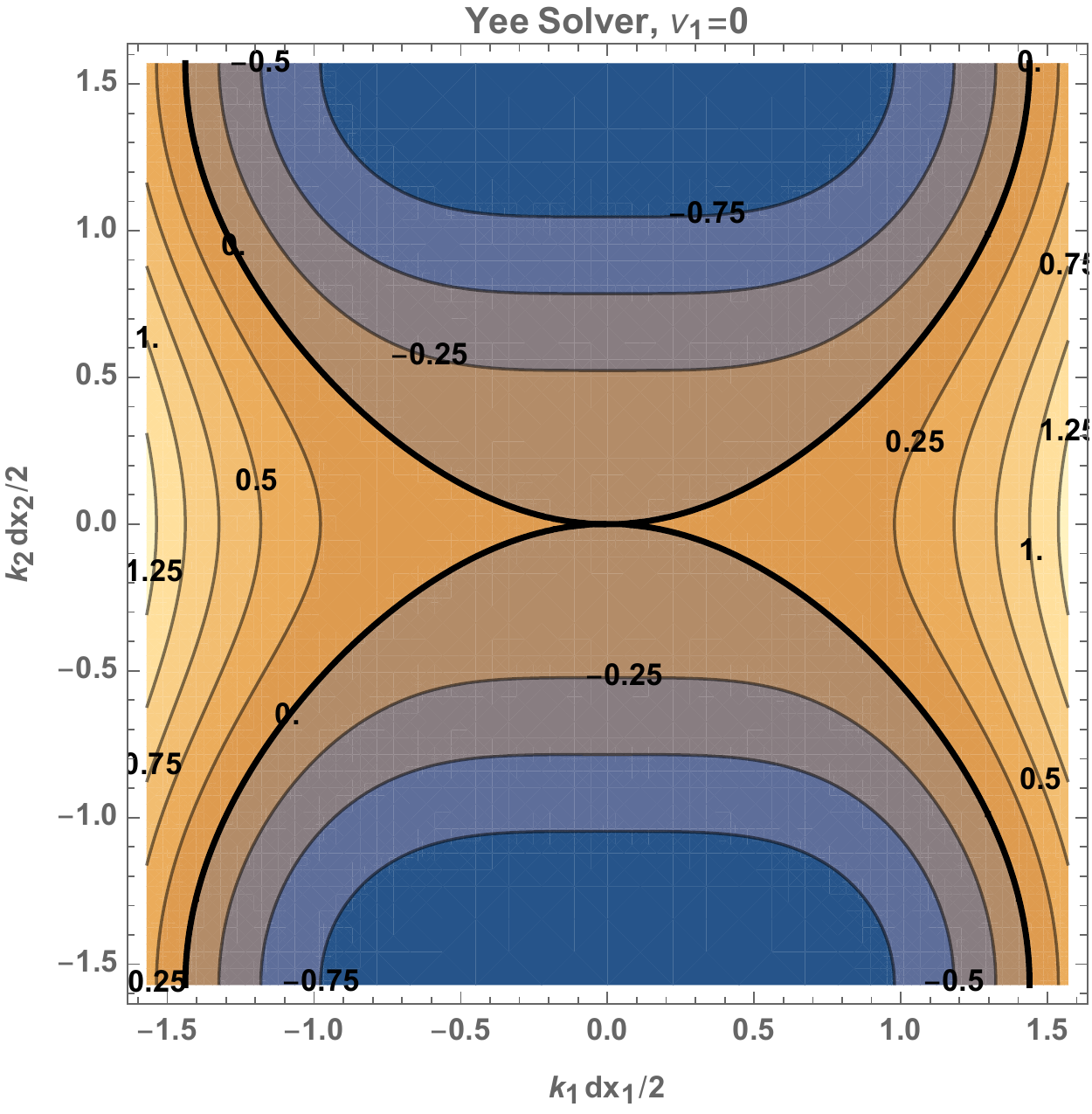}\includegraphics[width=0.5\textwidth]{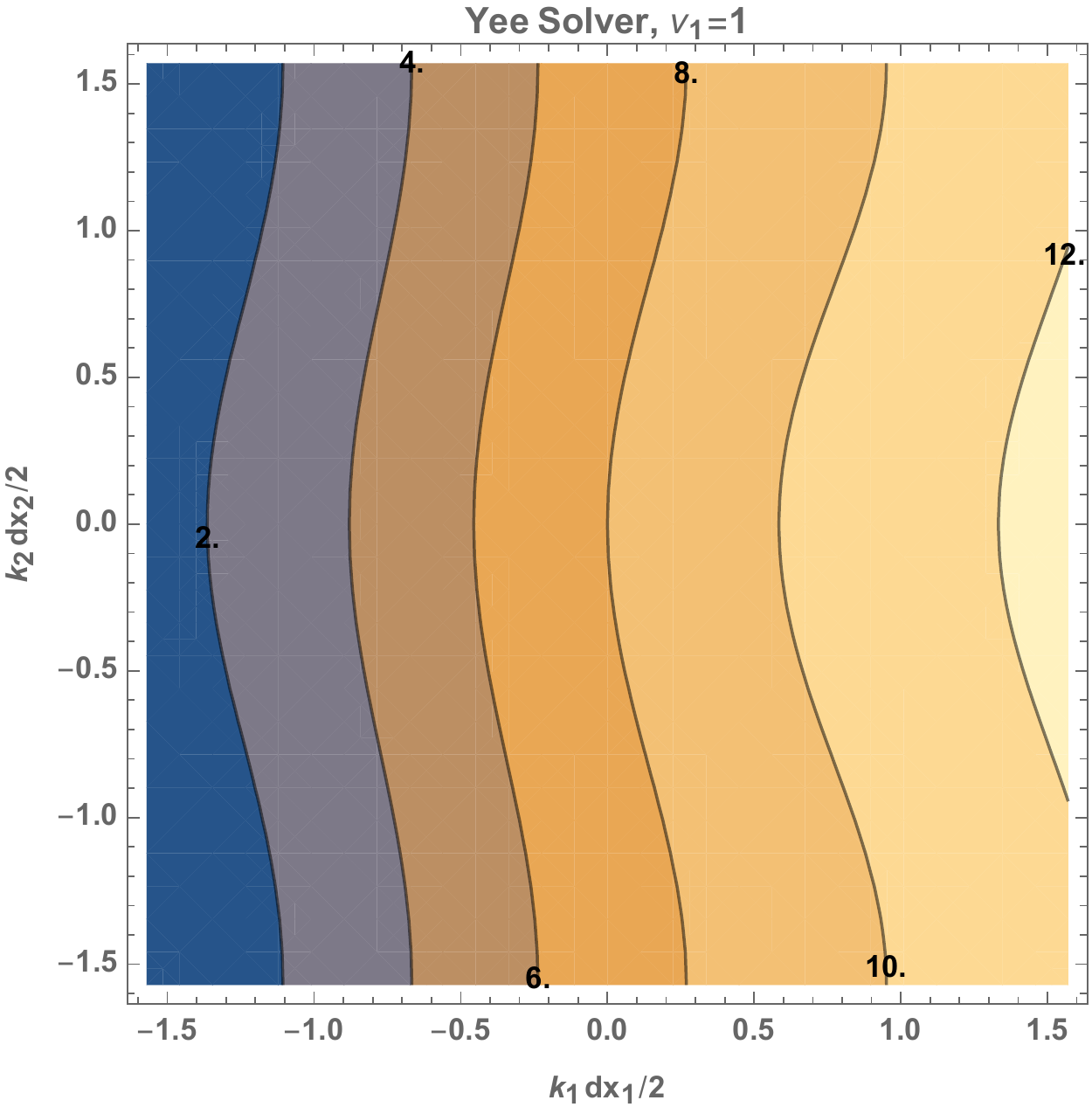}
\includegraphics[width=0.5\textwidth]{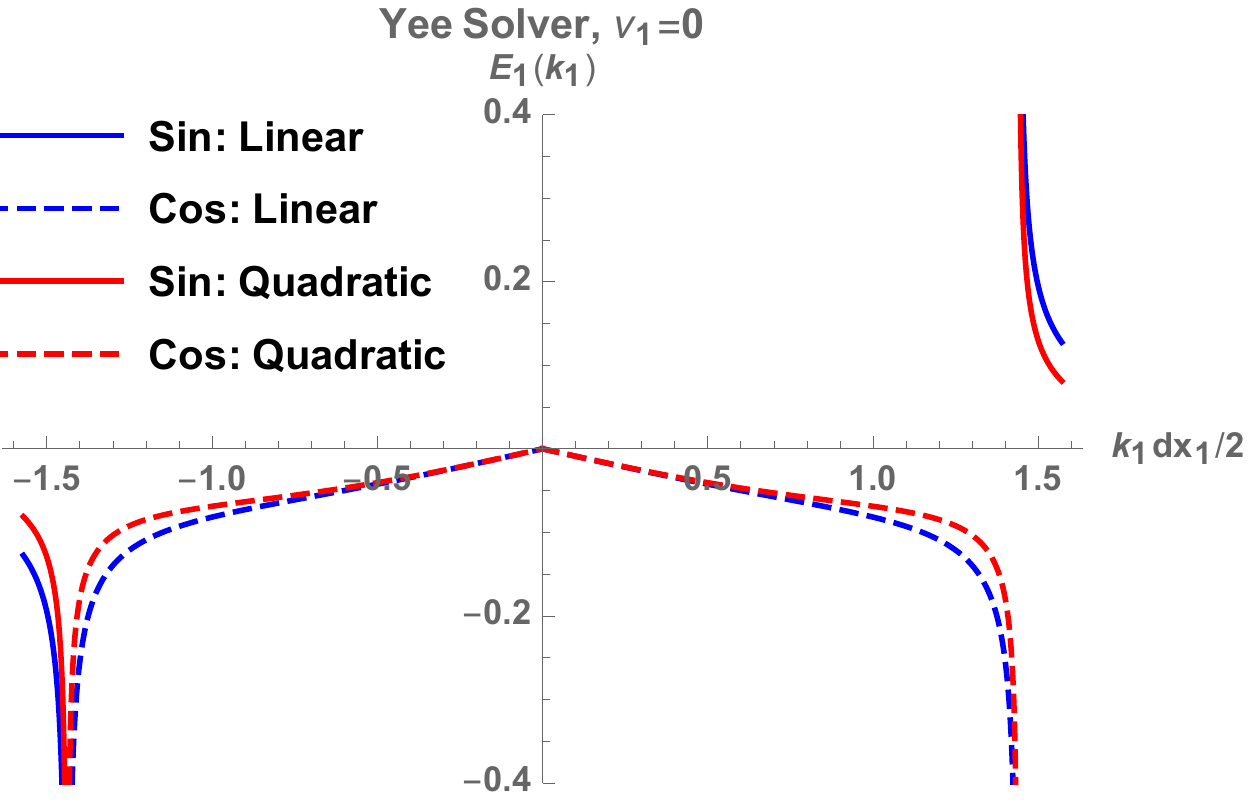}\includegraphics[width=0.5\textwidth]{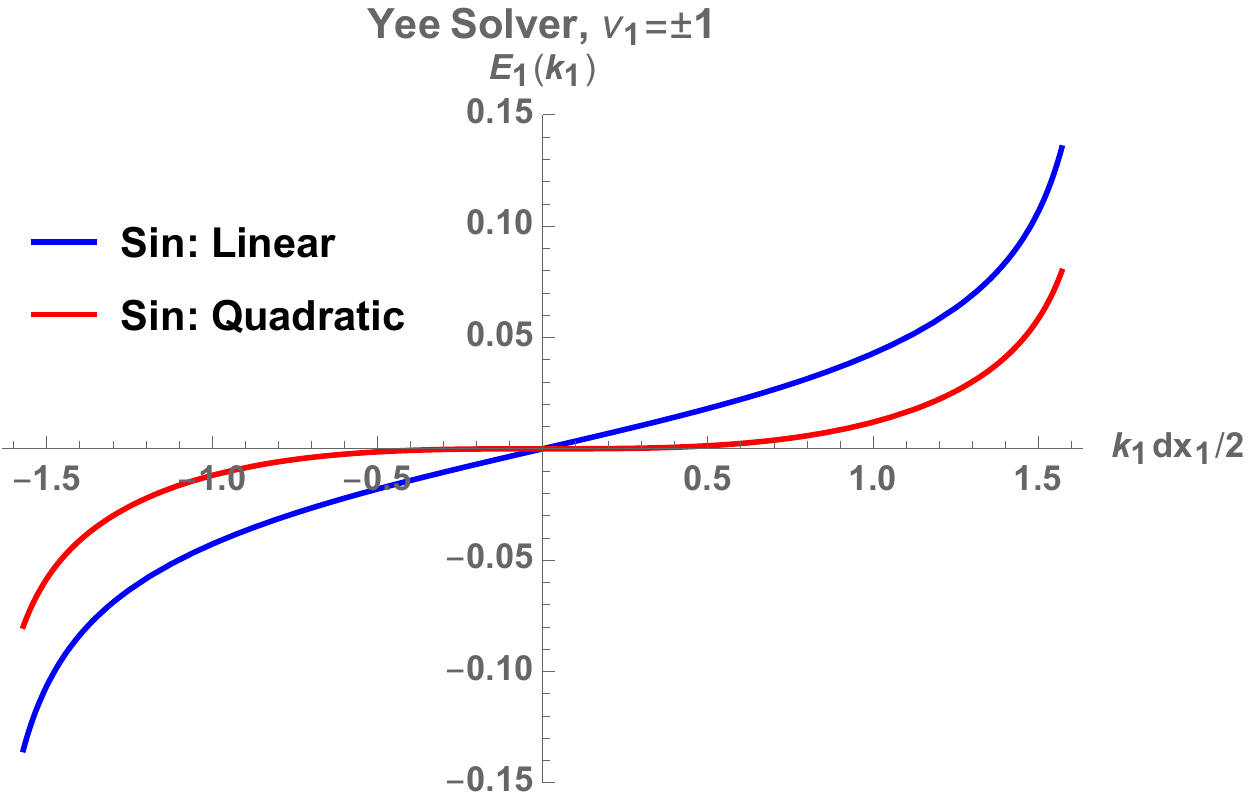}
\includegraphics[width=0.5\textwidth]{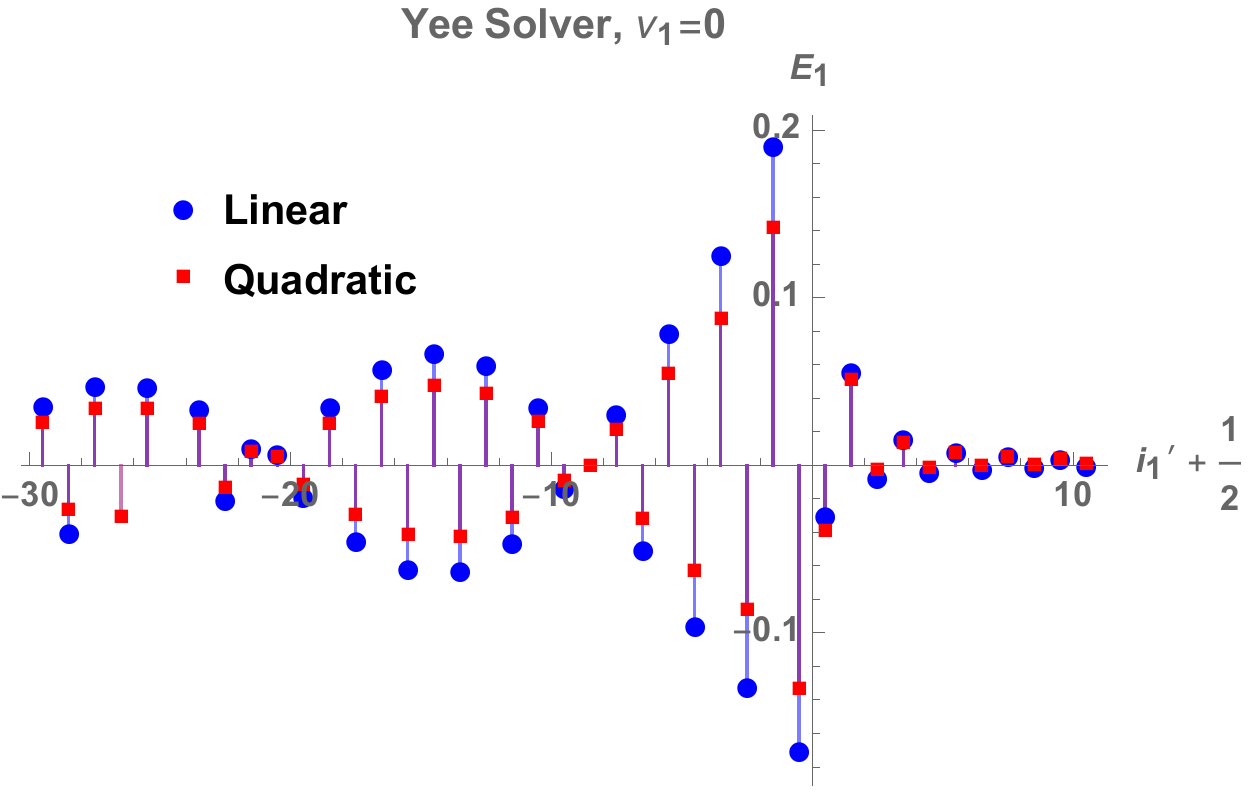}\includegraphics[width=0.5\textwidth]{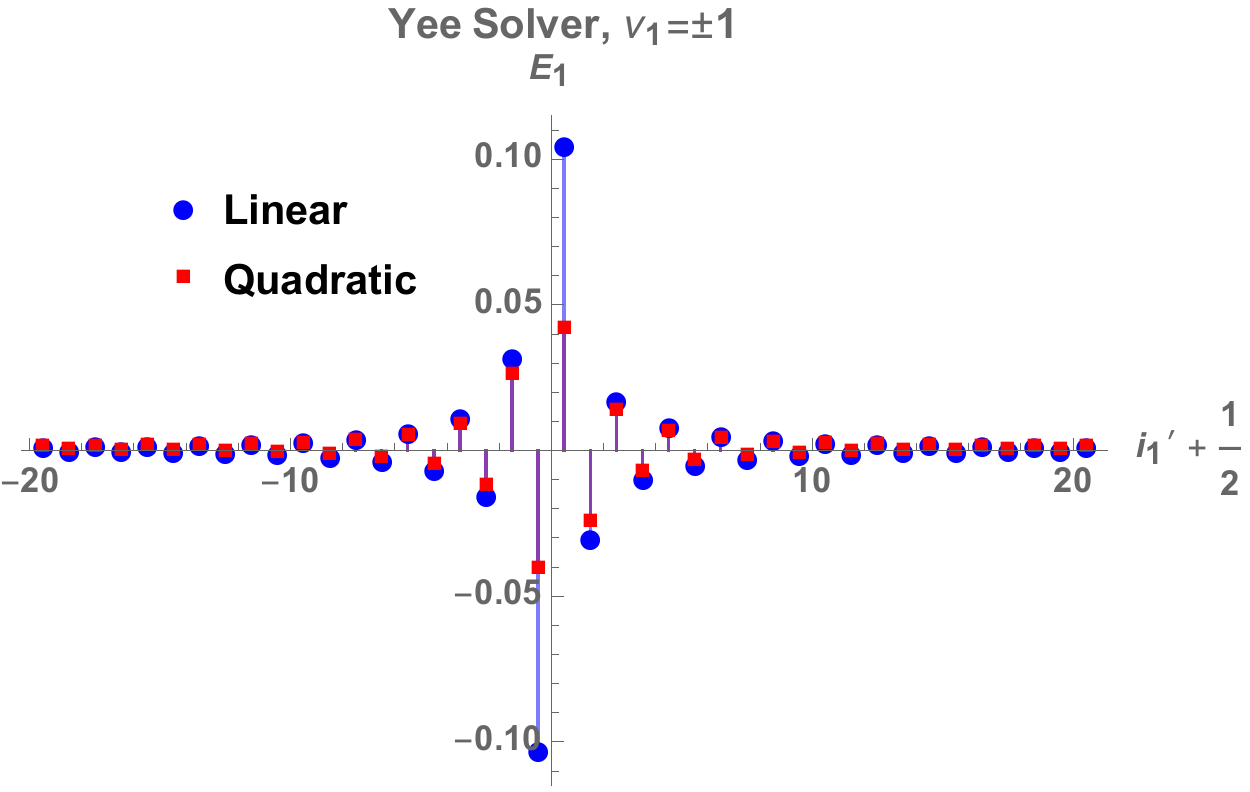}
\caption{ The value of $[k_1]_t^2-[k]_1^2 - [k]_2^2$ (upper row), the on-axis $E_1$ field in the $k_1$ space (middle row) and along the $x_1$ axis (bottom row) for the contribution from the fundamental and the first aliasing Brillouin zones for the Yee solver. Parameters: $dx_1=1, r\equiv \frac{dx_2}{dx_1}=1, \kappa\equiv \frac{dx_1}{dt}=4, q=1$.}
\label{fig: yee}
\end{center}
\end{figure}

For $\nu_1=0$ of the Yee solver, the explicit form of the denominator is $\frac{\mathrm{sin}^2(k_1 dt/2)}{(dt/2)^2} - \frac{\mathrm{sin}^2(k_1 dx_1/2)}{(dx_1/2)^2} - \frac{\mathrm{sin}^2(k_2 dx_2/2)}{(dx_2/2)^2}$. Depending on the values of $\kappa$ and $r$ this denominator will vanish for some $k_2$ for ranges of $k_1$. It can be shown that when $ r\leq \frac{2}{\sqrt{\pi^2 - 4}} \approx 0.83$ or $r> \frac{2}{\sqrt{\pi^2 - 4}}$,  and when $\kappa \leq \kappa_r$, where  $\kappa_r^2 \mathrm{sin}^2\left( \frac{\pi}{2\kappa_r}\right) = 1+r^{-2}$, a $k_2$ can be found for which the denominator vanishes. For these cases, the field structure will have a Cerenkov like radiation pattern. On the other hand,  when $r > \frac{2}{\sqrt{\pi^2 - 4}}$ and $\kappa > \kappa_r$, the fields have a contribution that is  Cerenkov like and another contribution that is space charge like. In this case, the denominator is positive definite for  $|k_1| > k_{1,r}$, while when $|k_1| \leq k_{1,r}$ the denominator can be zero for some $k_2$ where $k_{1,r}$  is defined by $\kappa^2\mathrm{sin}^2 \left( \frac{k_{1,r}dt}{2} \right) - \mathrm{sin}^2 \left( \frac{k_{1,r}dx_1}{2} \right) = r^{-2}$. We take the last condition, i.e., $r> \frac{2}{\sqrt{\pi^2 - 4} }$ and $\kappa > \kappa_r$, as an example to do the integrations over $k_2$ (details are given in Appendix D),
\begin{align}
 &-k_{1,r} \leq k_1\leq 0: 0 \leq [k_1]_t^2 - [k]_1^2 \leq \left( \frac{dx_2}{2} \right)^{-2}, \nonumber \\
 & \int_{-\frac{k_{g2}}{2}}^{\frac{k_{g2}}{2}}  \mathrm{d}k_2  \frac{ [ k_1 ]^2_t  - [k]_1^2} {  [ k_1 ]^2_t  - [k]_1^2 - [k]_2^2 }   =    i\pi\sqrt{ \frac{ [k_1]_t^2 - [ k]_1^2 } {  1 -  \left(\frac{dx_2}{2}\right)^2( [k_1 ]_t^2 - [k]_1^2 ) }   }	\nonumber 
 \end{align}
 \begin{align}
&0<k_1\leq k_{1,r}: 0 \leq [k_1]_t^2 - [k]_1^2 \leq \left( \frac{dx_2}{2} \right)^{-2}, \nonumber \\
&\int_{-\frac{k_{g2}}{2}}^{\frac{k_{g2}}{2}}  \mathrm{d}k_2  \frac{ [ k_1 ]^2_t  - [k]_1^2} {  [ k_1 ]^2_t  - [k]_1^2 - [k]_2^2 }  =   - i\pi\sqrt{ \frac{ [k_1]_t^2 - [ k]_1^2 } {  1 -  \left(\frac{dx_2}{2}\right)^2( [k_1 ]_t^2 - [k]_1^2 ) }   },  \nonumber 
\end{align}
\begin{align}
&|k_1| > k_{1,r} :   [k_1]_t^2 - [k]_1^2 > \left( \frac{dx_2}{2} \right)^{-2}, \nonumber \\
&\int_{-\frac{k_{g2}}{2}}^{\frac{k_{g2}}{2}}  \mathrm{d}k_2  \frac{ [ k_1 ]^2_t  - [k]_1^2} {  [ k_1 ]^2_t  - [k]_1^2 - [k]_2^2 }   =   \pi\sqrt{ - \frac{  [k_1]_t^2 -  [k]_1^2 } {  1  - \left(\frac{dx_2}{2}\right)^2(   [k_1 ]_t^2 - [k]_1^2  ) }   } \nonumber \\
\end{align}
where only the results for $i_2=0$ are shown.  Results for other grids in the transverse direction can be obtained through the appropriate integration. These fields have a more complicated form and they can be larger than the fields on axis for regions behind the particle. Note we must be careful to ensure the solutions satisfy causality and the Kramers-Kronig relations to get the correct sign of the integral when $-k_{1,r} \leq k_1< 0$ and $0<k_1\leq k_{1,r}$. Using the result in Eq. (16), the $E_1$ field from the contribution of the fundamental Brillouin zone can be obtained,
\begin{align}
&{E}^n_{1, i_1 , i_2=0 } (\nu_1=0)  \nonumber \\
&= - \frac{q}{2\pi}   \biggl(  \int_{0}^{k_{1,r}}  \mathrm{d}k_1   \frac{ S_1(k_{1})}{ [k]_1} \sqrt{ \frac{  [ k_1]^2_t - [k]_1^2 } {  1 -  \left(\frac{dx_2}{2}\right)^2(  [k_1 ]_t^2  - [k]_1^2 ) }   } \mathrm{cos}\left[ k_1dx_1 \left( i'_1  + \frac{1}{2} \right) \right]  \nonumber \\
&-  \int_{k_{1,r}}^{k_{g1}/2}    \mathrm{d}k_1   \frac{ S_1(k_{1})}{ [k]_1} \sqrt{ - \frac{  [ k_1]^2_t - [k]_1^2 } {  1 -  \left(\frac{dx_2}{2}\right)^2(  [k_1 ]_t^2  - [k]_1^2 ) }   } \mathrm{sin}\left[ k_1dx_1 \left( i'_1  + \frac{1}{2} \right) \right]  \biggl) \nonumber \\
&=  - \frac{q}{\pi dx_1}   \biggl( \int_{0}^{\hat{k}_{1,r}}  \mathrm{d}\hat{k}_1   \frac{ \hat{S}_1(\hat{k}_{1})}{ \mathrm{sin}\hat{k}_1} \sqrt{ \frac{ \kappa^2\mathrm{sin}^2 \frac{\hat{k}_1}{\kappa}  - \mathrm{sin}^2\hat{k}_1} {  1 - r^2 \left(  \kappa^2\mathrm{sin}^2 \frac{\hat{k}_1}{\kappa}  - \mathrm{sin}^2\hat{k}_1 \right)  }   } \mathrm{cos}\left[ \hat{k}_1( 2i'_1  + 1 ) \right]  \nonumber \\
&- \int_{\hat{k}_{1,r}}^{\frac{\pi}{2}}    \mathrm{d}\hat{k}_1   \frac{ \hat{S}_1(\hat{k}_{1})}{ \mathrm{sin}\hat{k}_1} \sqrt{  - \frac{ \kappa^2\mathrm{sin}^2 \frac{\hat{k}_1}{\kappa} - \mathrm{sin}^2\hat{k}_1  } {  1 - r^2 \left(  \kappa^2\mathrm{sin}^2 \frac{\hat{k}_1}{\kappa} -  \mathrm{sin}^2\hat{k}_1  \right)  }    } \mathrm{sin}\left[ \hat{k}_1( 2i'_1  + 1 ) \right]  \biggl) \nonumber \\
\end{align}
where $\hat{k}_1 = \frac{k_1 dx_1}{2}$. The numerical results are shown in the left column of Fig. \ref{fig: yee}. The Cerenkov radiation pattern dominates the field contributions from the fundamental zone. As a result, the on-axis $E_1$ field is large and it extends 
 behind the particle (the fields behind the particle are in fact larger off axis). These Cerenkov fields do not extend much in front of the particle. The fields in front of the particle are dominated by the space charge like fields. The use of  higher-order particle shapes can reduce the high $k_1$ spectral components from the fundamental Brillouin zone and thus decrease the unphysical fields.  

We next discuss the contributions to the unphysical fields from aliasing, i.e., from the higher order Brillouin zones. For the $\nu_1=\pm1$ zones of the Yee solver, the value of the denominator is now $\frac{\mathrm{sin}^2[ (k_1\pm k_{g1}) dt/2 ]}{(dt/2)^2} - \frac{\mathrm{sin}^2(k_1 dx_1/2)}{(dx_1/2)^2} - \frac{\mathrm{sin}^2(k_2 dx_2/2)}{(dx_2/2)^2}$ which becomes more complicated. Here, we do not list all the possible regions of $r$ and $\kappa$ space, but take $r=1, \kappa\geq2$ as an example. For other values of $r$ and $\kappa$, the reader can analyze it similarly. When $r=1, \kappa\geq2$, it can be shown that $[k_1\pm k_{g1}]_t^2 - [k]_1^2 - [k]_2^2 >0 $, i.e., the integration function has no singularities in the entire integration region.  Thus the field is all space charge like and can be written as
\begin{align}
&E^n_{1,i_1,i_2=0} (\nu_1= -1) + E^n_{1,i_1,i_2} (\nu_1= 1)\nonumber \\
&= - \frac{q}{4\pi}  \int_{-\frac{k_{g1}}{2}}^{\frac{k_{g1}}{2}}  \mathrm{d}k_1   \frac{i}{ [k]_1}  \biggl[ S_1(k_{1} - k_{g1}) \sqrt{ - \frac{ [ k_1 - k_{g1}]^2_t  - [k]_1^2 } {  1 - \left( \frac{dx_2}{2}\right)^2(  [k_1 - k_{g1}]_t^2 - [k]_1^2 ) }   }   \nonumber \\ 
&+ S_1(k_{1} + k_{g1})\sqrt{ -\frac{ [ k_1+k_{g1}]^2_t -  [k]_1^2 } {  1 - \left( \frac{dx_2}{2}\right)^2 (  [k_1  + k_{g1}]_t^2 - [k]_1^2 ) }   }  \biggr] \mathrm{exp}\left[ i k_1dx_1 \left( i'_1  + \frac{1}{2} \right) \right]  \nonumber \\
&=  \frac{q}{\pi dx_1} \int_{0}^{\frac{\pi}{2}}  \mathrm{d}\hat{k}_1   \biggl[    \frac{\hat{S}_1(\hat{k}_{1} - \pi)}{\mathrm{sin}\hat{k}_1}   \sqrt{  - \frac{  \kappa^2 \mathrm{sin}^2\left( \frac{ \hat{k}_1 - \pi}{\kappa}\right)  - \mathrm{sin}^2\hat{k}_1  } {  1 - r^2 \left[  \kappa^2 \mathrm{sin}^2\left( \frac{ \hat{k}_1 - \pi}{\kappa}\right)  -\mathrm{sin}^2\hat{k}_1 \right] }   }  \nonumber \\
&+ \frac{\hat{S}_1(\hat{k}_{1}+ \pi)}{\mathrm{sin}\hat{k}_1}   \sqrt{ - \frac{  \kappa^2 \mathrm{sin}^2 \frac{ \hat{k}_1 + \pi}{\kappa} - \mathrm{sin}^2\hat{k}_1  } {  1 - r^2 \left(  \kappa^2 \mathrm{sin}^2 \frac{ \hat{k}_1 + \pi}{\kappa}  -\mathrm{sin}^2\hat{k}_1 \right) }   }    \biggr] \mathrm{sin}\left[ \hat{k}_1( 2i'_1  + 1) \right] 
\end{align}
The numerical results are shown in the right column of Fig. \ref{fig: yee}. For the particular parameters studied here, a pure space charge pattern exist thus the $E_1$ field is antisymmetric in $i_1$ about the particle; hence there are no self-forces from these zones. The amplitude of the on-axis $E_1$ is reduced by a factor of around 2 by using a quadratic particle shape instead of a linear shape. 

\begin{figure}[htbp]
\begin{center}
\includegraphics[width=0.5\textwidth]{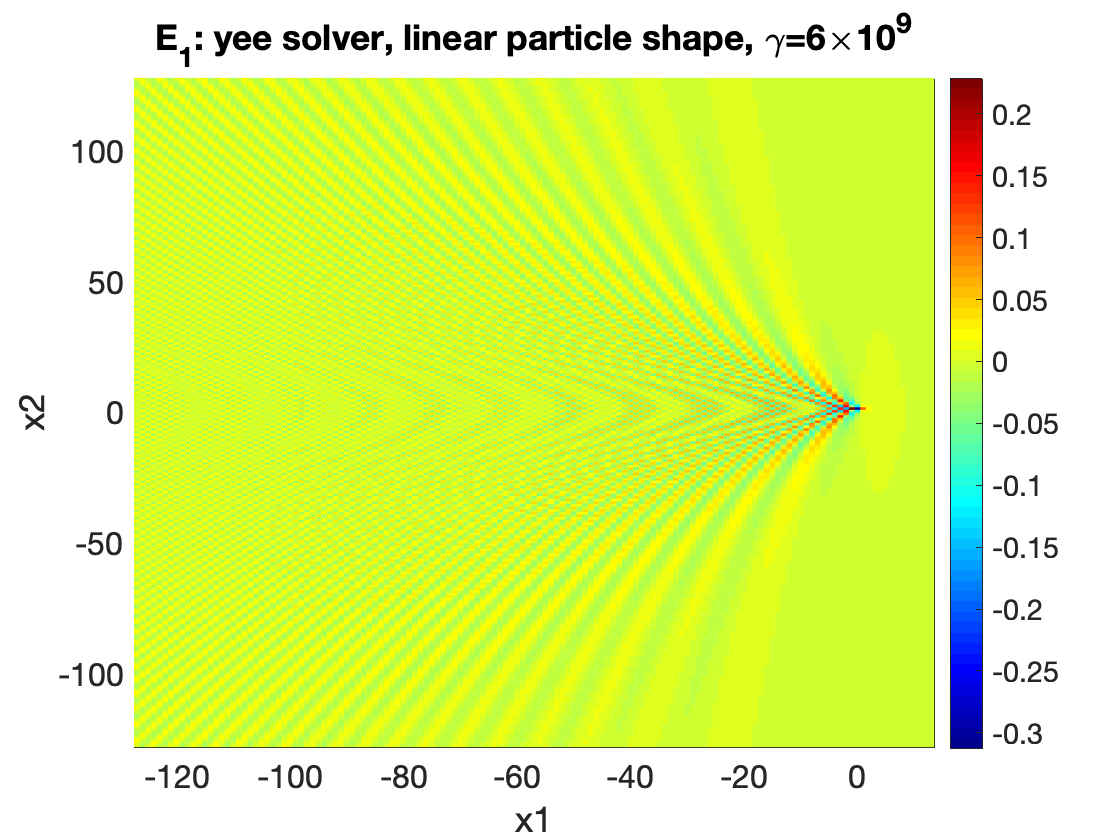}\includegraphics[width=0.5\textwidth]{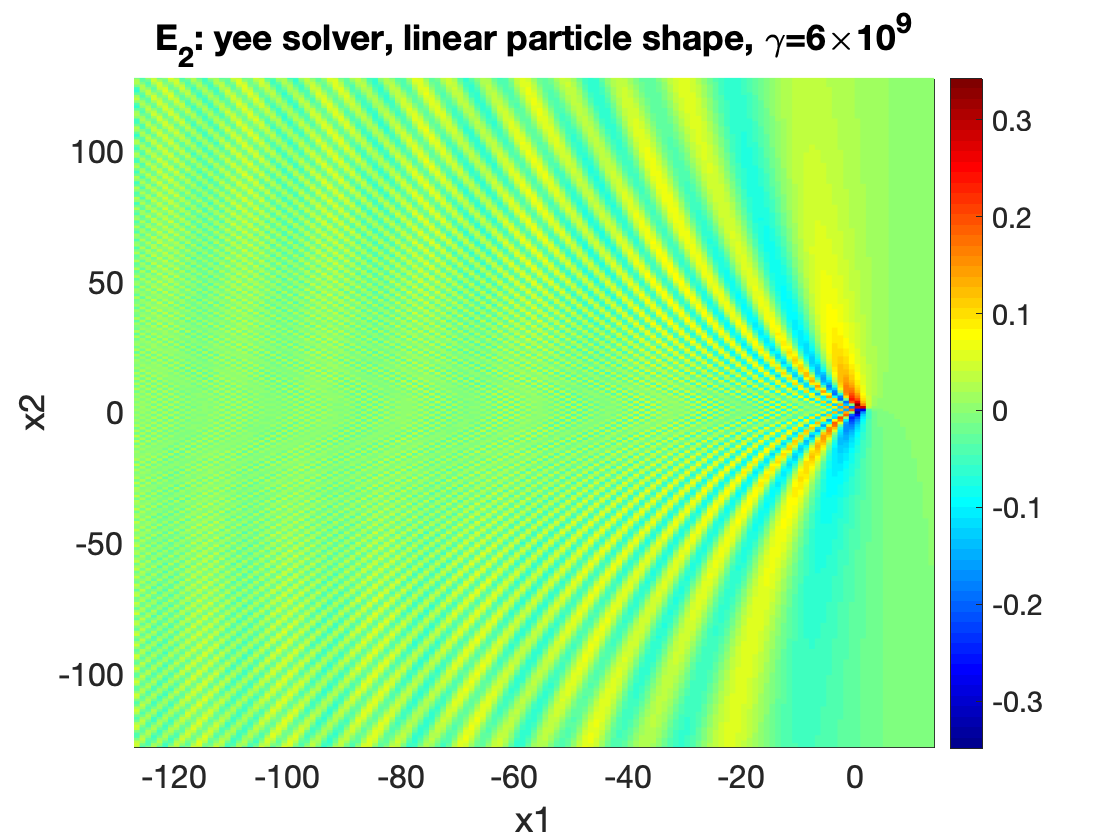}
\includegraphics[width=0.5\textwidth]{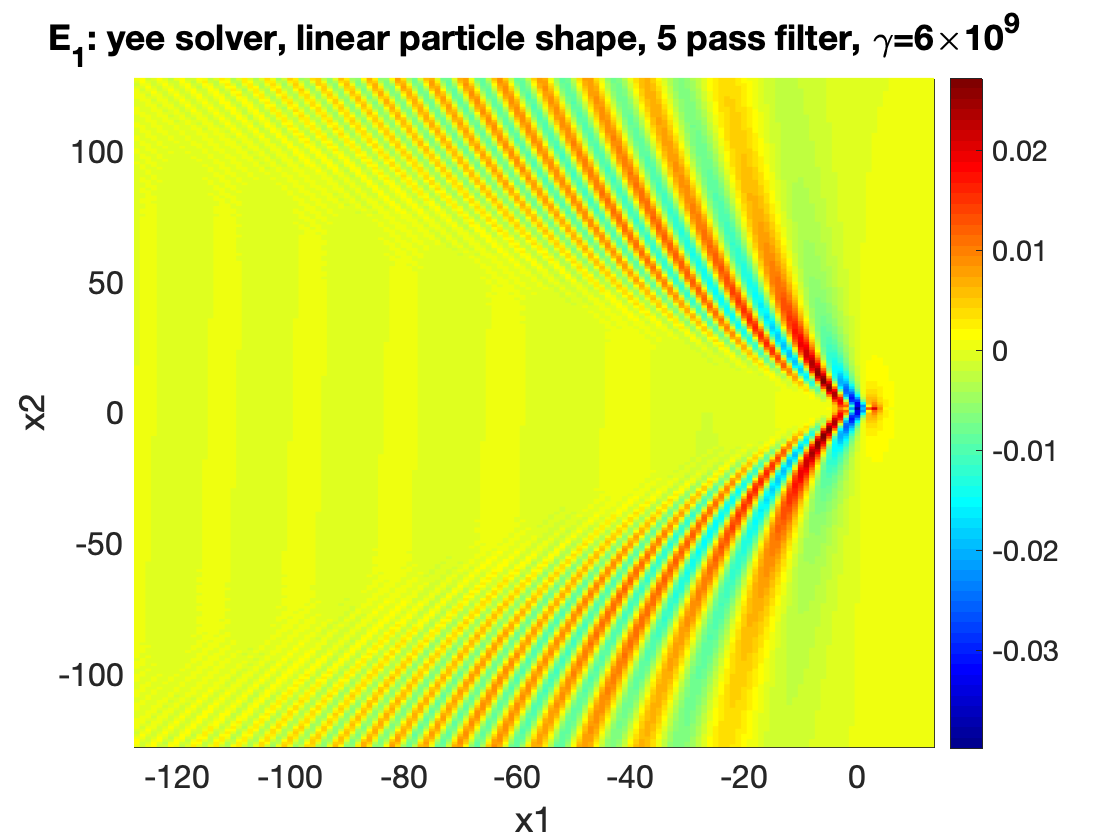}\includegraphics[width=0.5\textwidth]{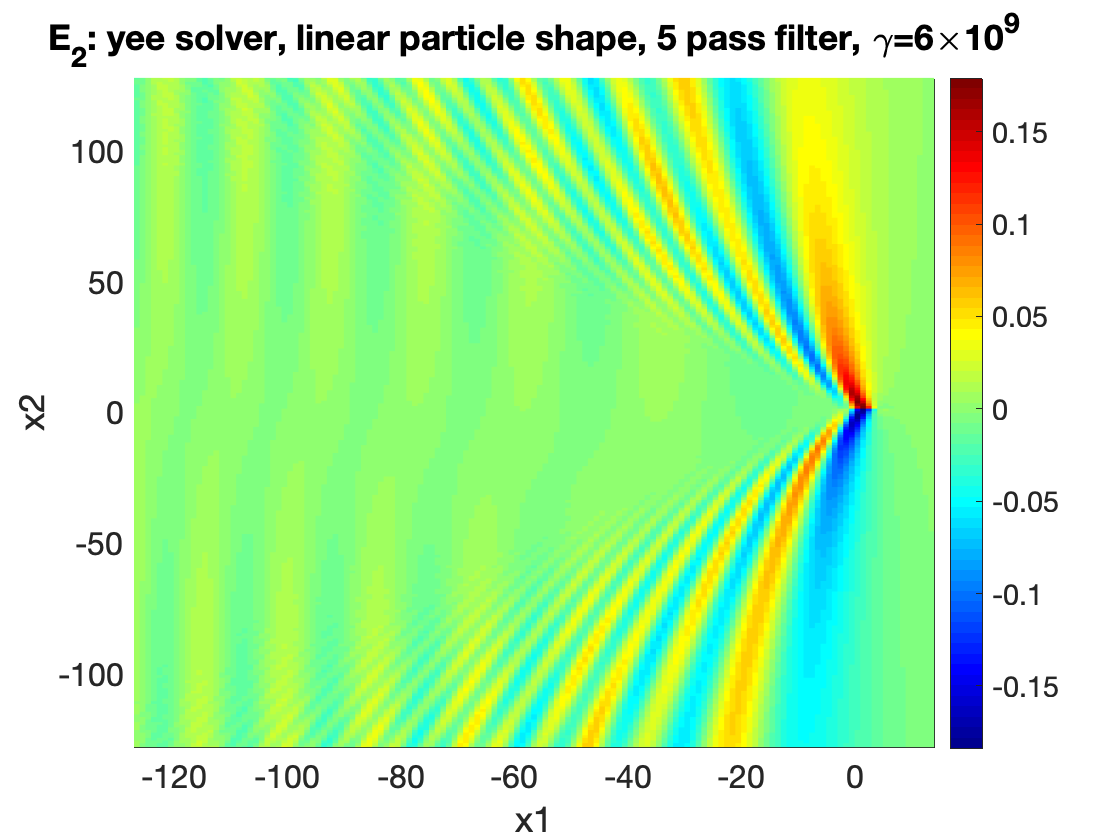}
\includegraphics[width=0.5\textwidth]{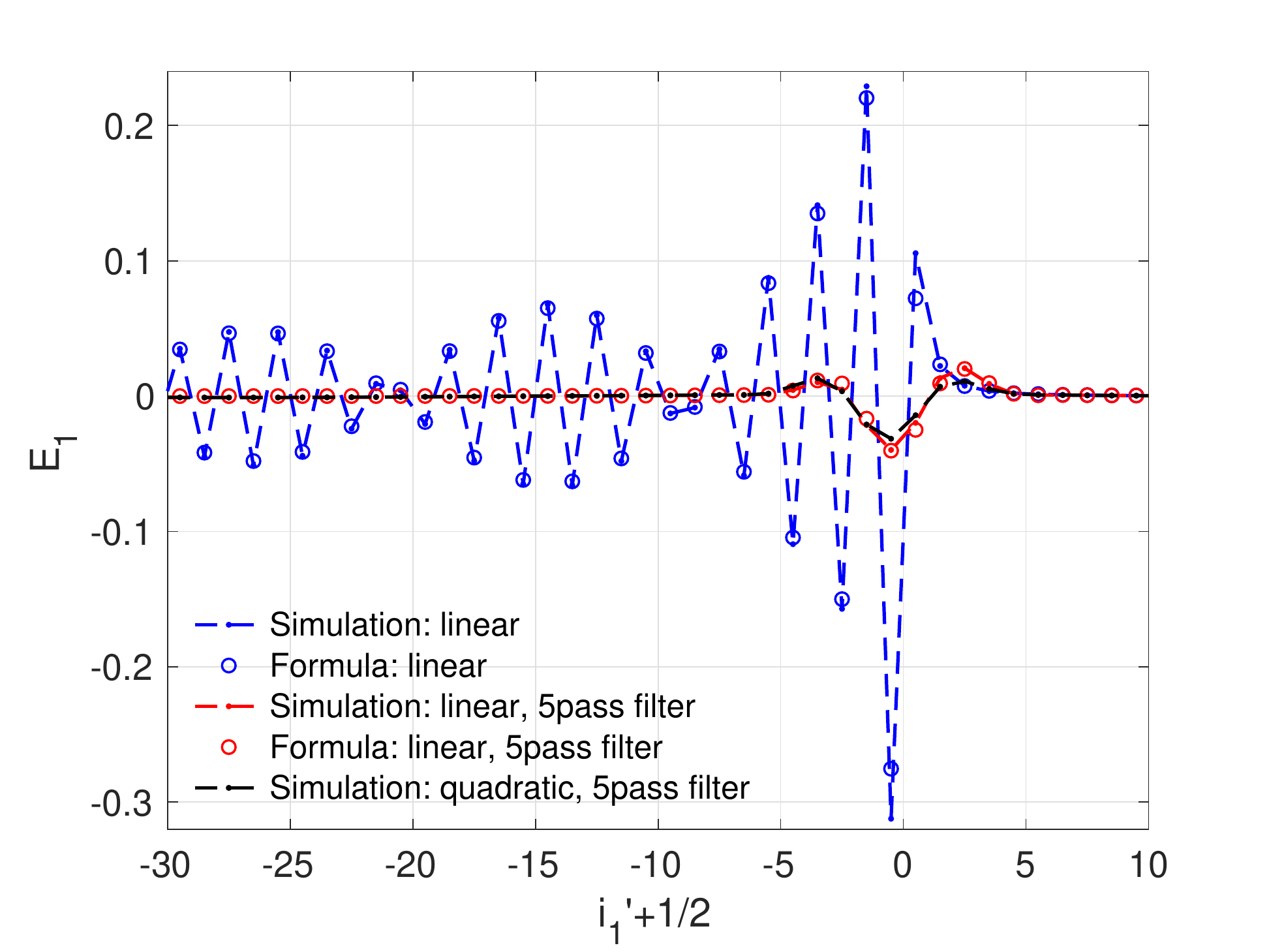}\includegraphics[width=0.5\textwidth]{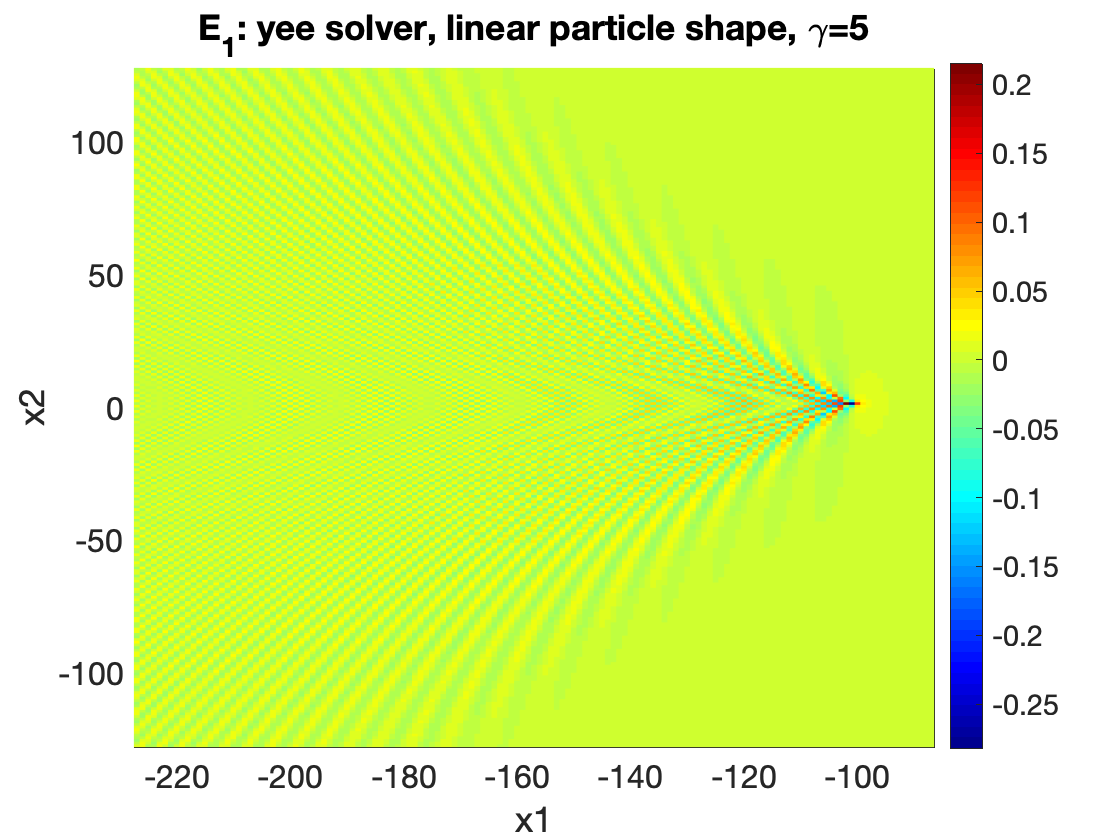}
\caption{The electric fields $E_1$ and $E_2$ of a free-streaming particle from OSIRIS simulations with the Yee solver after 20000 time steps. Parameters: $dx_1=1, r\equiv \frac{dx_2}{dx_1}=1, \kappa\equiv \frac{dx_1}{dt}=4, q=1$, other simulation parameters are shown in each subplot. }
\label{fig: yee sim}
\end{center}
\end{figure}

Simulation results from OSIRIS of a free streaming relativistic particle  and their comparison with the formulas are shown in Fig. \ref{fig: yee sim}. The analytical result, Fig. \ref{fig: yee}, is confined to the location of the particle. As can be seen, the simulation fields are large and extend far behind the particle, thus we can see that the  transverse and longitudinal fields are dominated by the numerical fields. A 5 pass filter can reduce the high wave number components, however the remaining unphysical fields are still unacceptably large, e.g., $E_1\frac{dx_1}{q} \sim 10^{-2}$. The use of quadratic particle shapes also does not reduce the numerical fields to acceptable levels even when combined with the 5 pass filter \cite{5passfilter}. The comparisons for the on-axis $E_1$ field between the simulation results and the analytical expressions for PIC codes are shown in the bottom left of Fig. \ref{fig: yee sim}. Excellent agreement is seen. The small deviations close to the particle position may be due to the contribution from higher Brillouin zones ($|\nu_1| \geq 2$) which are not included in the formulas. The $E_1$ field for a particle with $\gamma=5$ is shown in the bottom right of Fig. \ref{fig: yee sim} which is similar to the field from a particle with $\gamma=6\times 10^9$. This is because the fields are dominated by the numerical issues not by the particle's energy ($\gamma$ factor) as we discussed earlier.

The particle loses energy through its numerical Cerenkov radiation in the fundamental Brillouin zone. As a result the particle will decelerate (there is still energy conservation between kinetic and field energy). This energy loss appears as a "self-force" and can be obtained by integrating the axial field  over the particle, 
\begin{align}
F_1&=  \int_{-\frac{k_{g1}}{2}}^{\frac{k_{g1}}{2}} \mathrm{d}k_1 q S(k_1) \tilde{E}_1(k_1)\nonumber \\
&\approx - \frac{ q^2}{\pi dx_1}   \int_{0}^{\hat{k}_{1,r}}  \mathrm{d}\hat{k}_1   \frac{ \hat{S}^2_1(\hat{k}_{1})}{ \mathrm{sin}\hat{k}_1} \sqrt{ \frac{ \kappa^2\mathrm{sin}^2\frac{\hat{k}_1}{\kappa}  - \mathrm{sin}^2\hat{k}_1} {  1 - r^2 \left[  \kappa^2\mathrm{sin}^2\frac{\hat{k}_1}{\kappa}  - \mathrm{sin}^2\hat{k}_1 \right]  }   }
\end{align}
where only the contributions from $\nu_1=0,\pm1$ are considered. This self-force and the energy loss through a given distance are inversely proportional to $dx_1$ when $\kappa$ and $r$ are fixed. For the parameters examined in Fig. \ref{fig: yee sim}, the expression for the self-force can be integrated numerically to obtain $F_1 \approx -0.15 q^2/dx_1$ for a linear particle shape, $-0.10 q^2/dx_1$ for a quadratic shape,  and $-0.031q^2/dx_1$ for a quadratic shape and a 5 pass filter. 

\subsection{The fields with the spectral solver}
In a spectral solver the fields are advanced in $\bm{k}$ space and it generally has improved dispersion properties. Some refer to such a scheme as pseudo spectral because 
a grid is used. The phase velocity of the EM waves for a spectral solver is faster than the speed of light in vacuum, thus such a solver is not as susceptible to numerical Cerenkov effects. It recently it was shown to suppress the NCI.  For a spectral solver, we have
\begin{align}
[\omega]_t= \frac{\mathrm{sin}(\omega \frac{dt}{2})}{\left(\frac{dt}{2}\right)}, [k]_{E1} = [k]_{B1} = k_1,  [k]_{E2} = [k]_{B2} = k_2
\end{align}

Substituting these expressions into Eq. (\ref{eq: E12D}) leads to the expression for the $E_1$ field in the fundamental zone, 
\begin{align}
{E}^n_{1, i_1 , i_2=0 } (\nu_1=0)  &=   -\frac{q}{(2\pi)^2} \int_{-\frac{k_{g1}}{2}}^{\frac{k_{g1}}{2}} \int_{-\frac{k_{g2}}{2}}^{\frac{k_{g2}}{2}} \mathrm{d}k_1 \mathrm{d}k_2  \frac{i  S(k_{1})}{k_1}\frac{ [ k_1]^2_t - k_1^2 } {  [ k_1]^2_t  - k_1^2 - k_2^2 }  \mathrm{exp}\left[ i k_1 \left(i'_1+\frac{1}{2}\right) dx_1 \right] 
\end{align}
In order to compare with the simulation results from OSIRIS, staggered grids are used here. The results are similar with non-staggered grids as used in other PIC codes with a spectral solver. 

\begin{figure}[htbp]
\begin{center}
\includegraphics[width=0.5\textwidth]{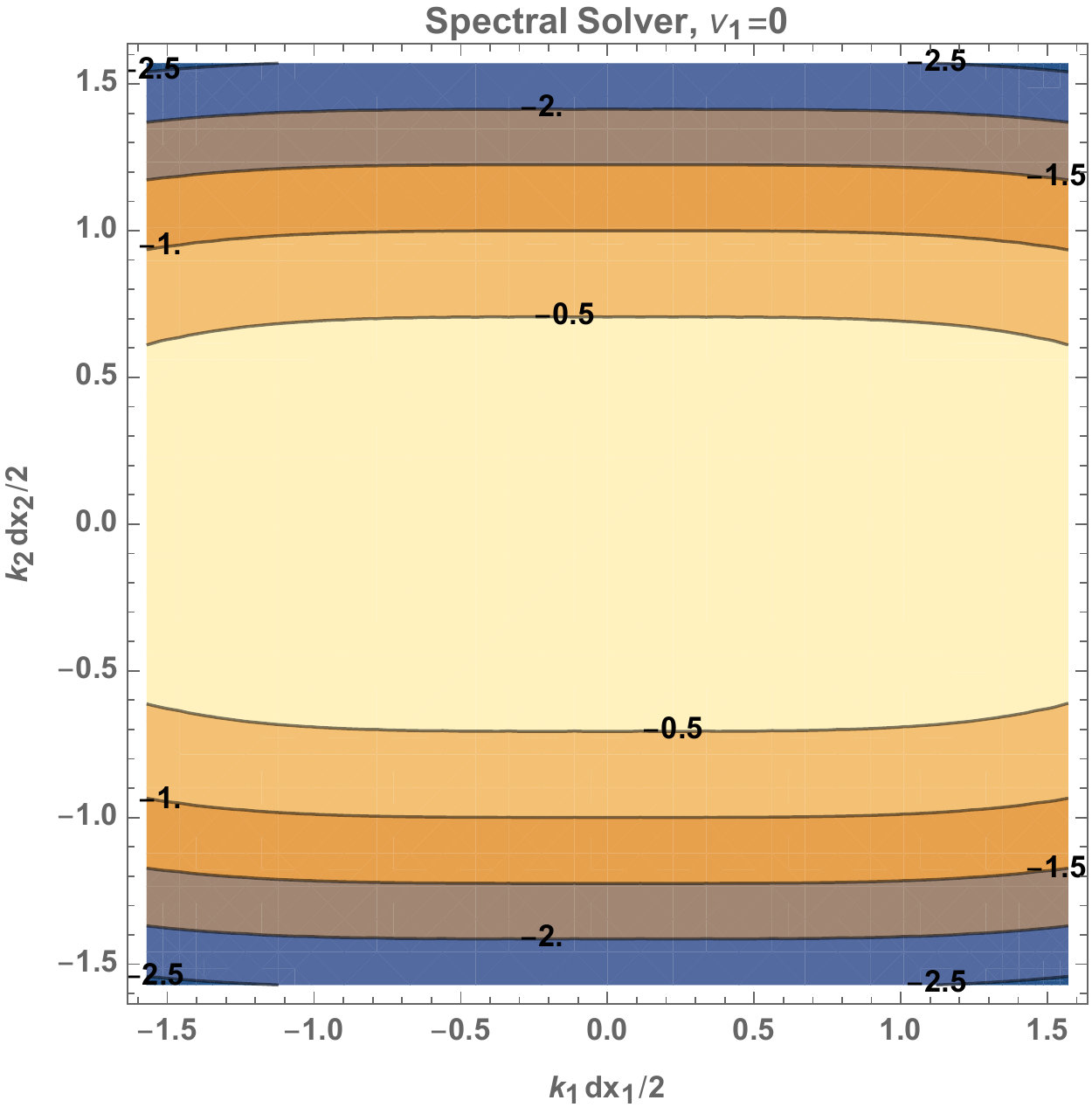}\includegraphics[width=0.5\textwidth]{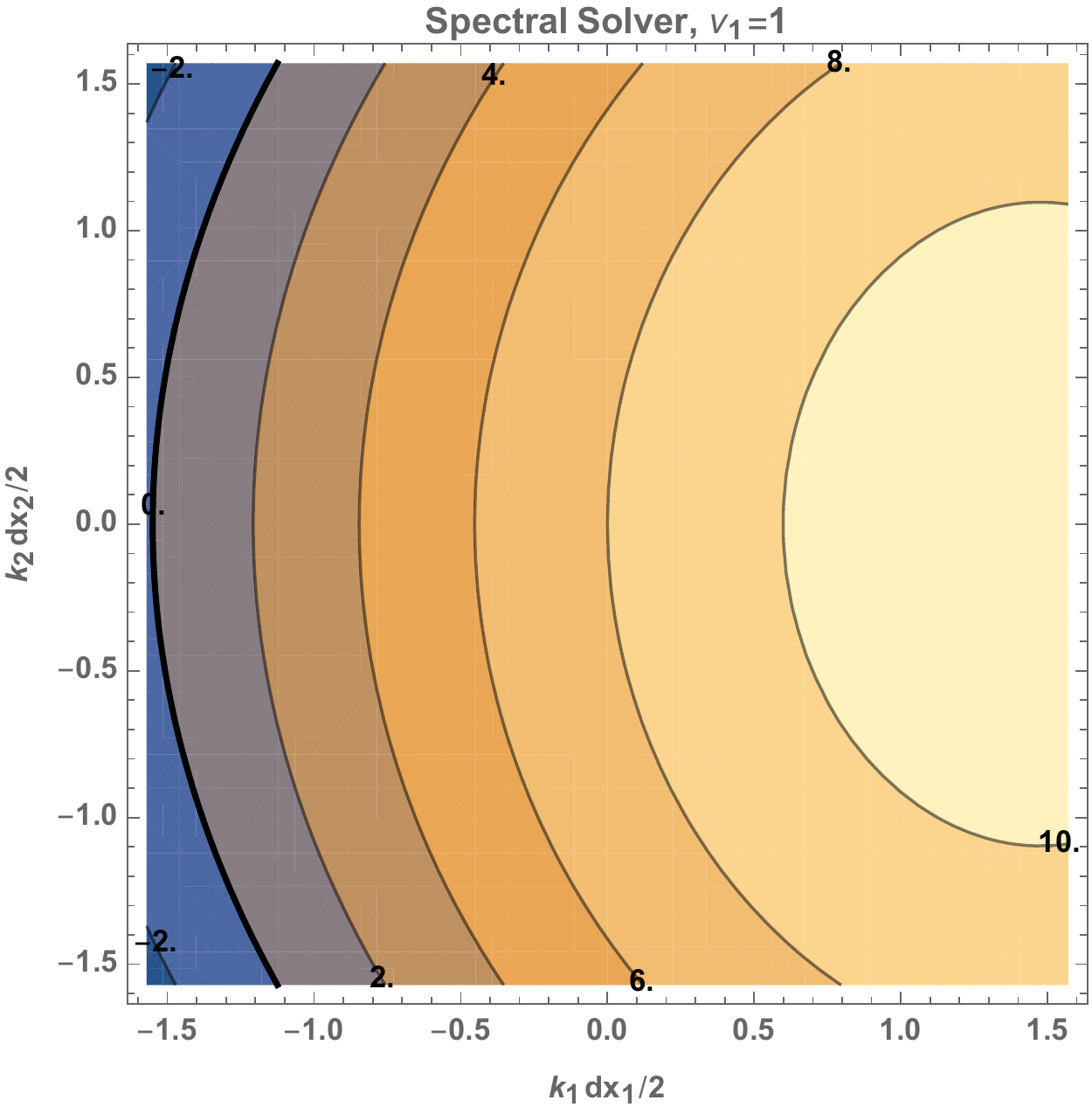}
\includegraphics[width=0.5\textwidth]{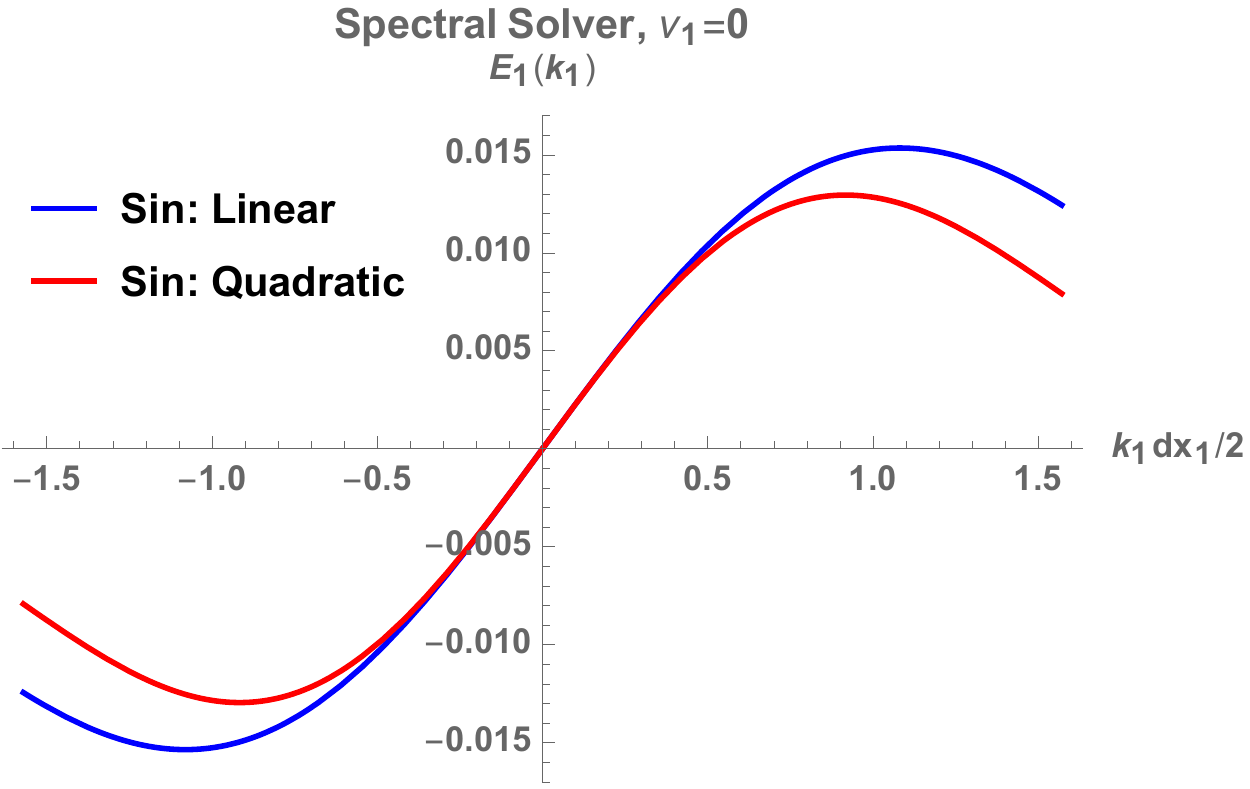}\includegraphics[width=0.5\textwidth]{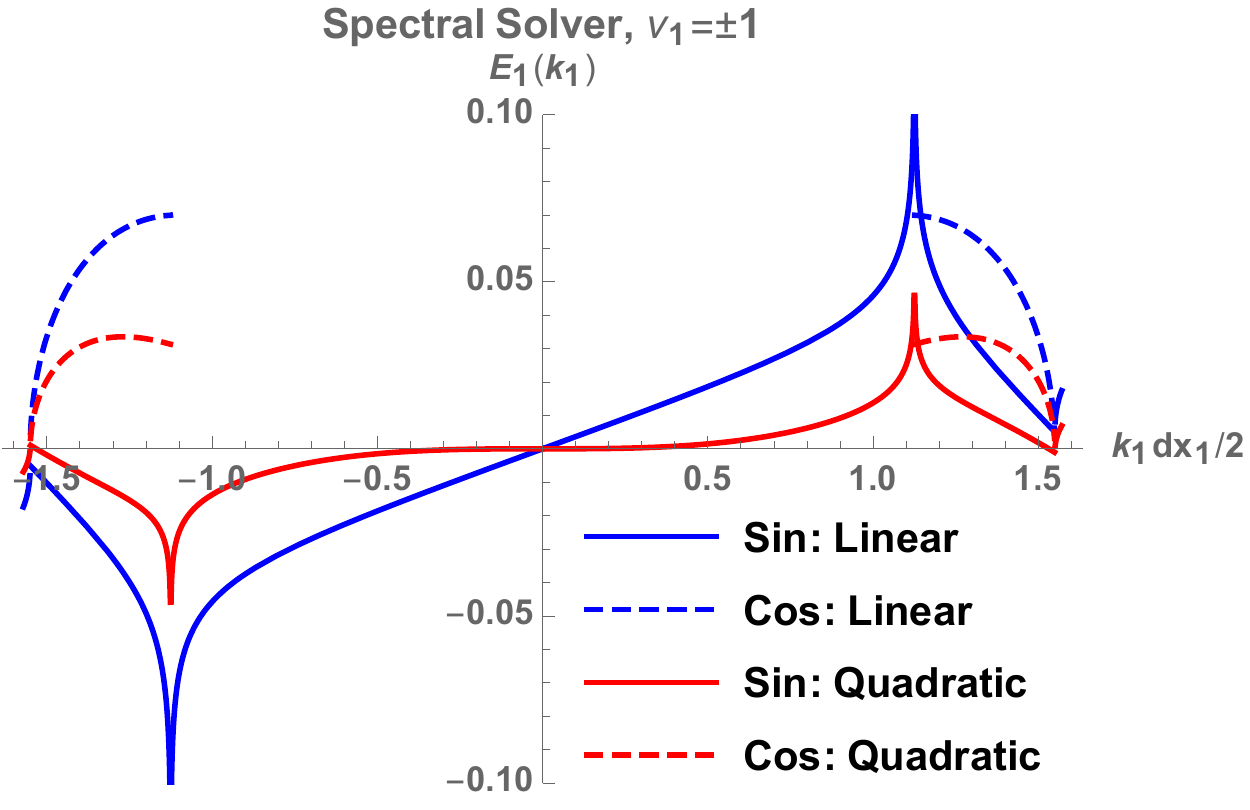}
\includegraphics[width=0.5\textwidth]{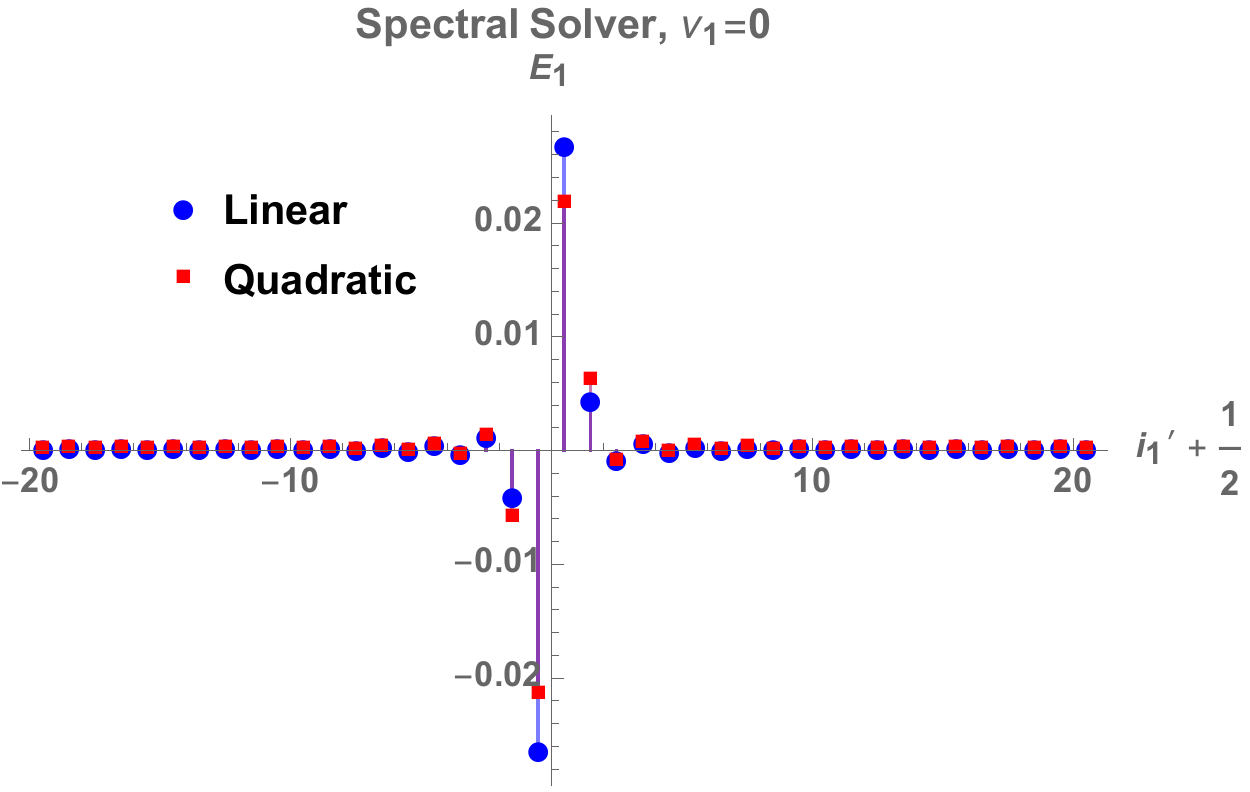}\includegraphics[width=0.5\textwidth]{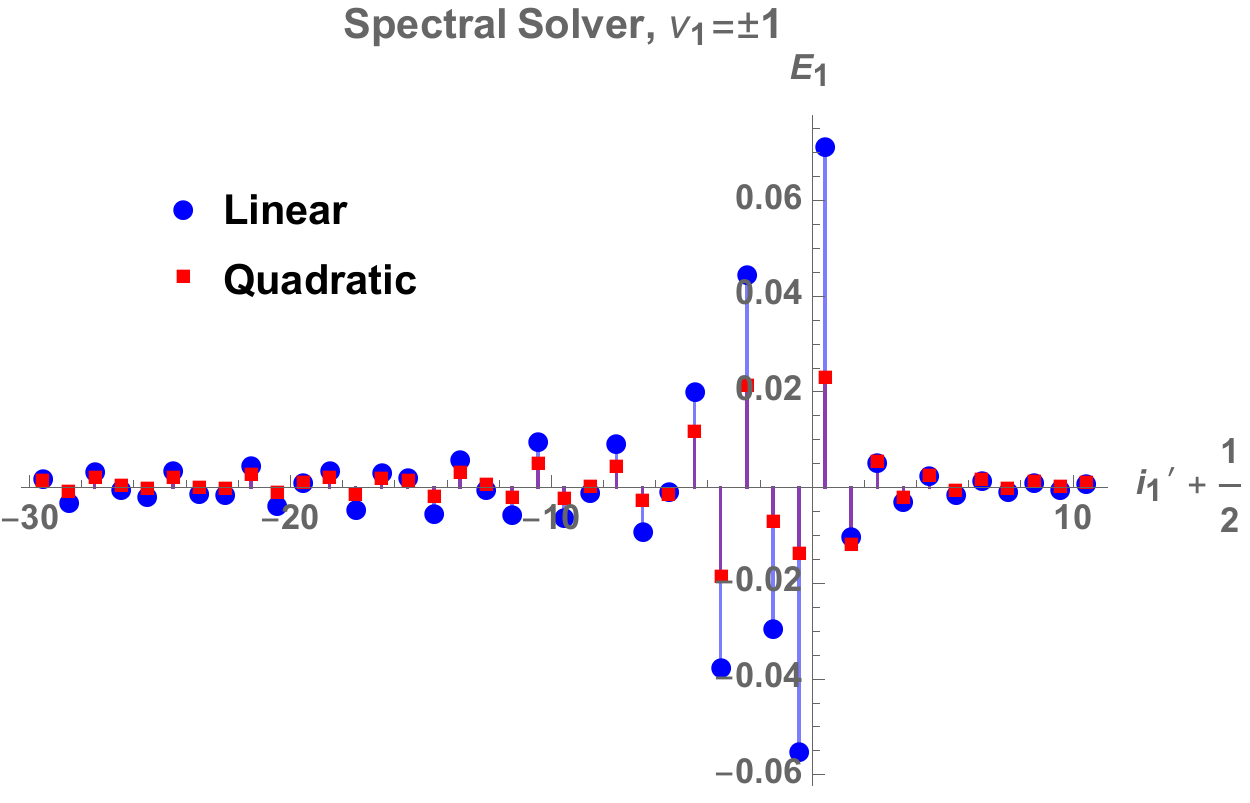}
\caption{ The value of $[k_1]_t^2-[k]_1^2 - [k]_2^2$ (upper row), the on-axis $E_1$ field in the $k_1$ space (middle row) and along the $x_1$ axis (bottom row) for the contribution from the fundamental and the first aliasing Brillouin zones for the spectral solver. Parameters: $dx_1=1, r\equiv \frac{dx_2}{dx_1}=1, \kappa\equiv \frac{dx_1}{dt}=4, q=1$.}
\label{fig: spe}
\end{center}
\end{figure}

For all possible $\kappa$ and $r$, the denominator of the integrand $[ k_1]^2_t  - k_1^2 - k_2^2 \leq 0$; thus there are no Cerenkov like fields in the fundamental zone, the numerical fields are all space charge like,
\begin{align}
&{E}^n_{1, i_1 , i_2=0 } (\nu_1=0) \nonumber \\
& =   -\frac{q}{(2\pi)^2} \int_{-\frac{k_{g1}}{2}}^{\frac{k_{g1}}{2}} \mathrm{d}k_1   \frac{ i S(k_{1})}{ k_1}  2\sqrt{ k_1^2 -  [k_1]_t^2 } \mathrm{tan}^{-1}\left( \frac{k_{g2}}{2\sqrt{k_1^2 - [k_1]_t^2 } }\right)   \mathrm{exp}\left[ i k_1 \left(i'_1+\frac{1}{2}\right) dx_1 \right] 	\nonumber \\
&= \frac{2q}{ \pi^2 dx_1}\int_{0}^{\frac{\pi}{2}} \mathrm{d} \hat{k}_1  \frac{ \hat{S}(\hat{k}_{1})}{  \hat{k}_1 } \sqrt{ \hat{k}_1^2 -  \kappa^2 \mathrm{sin}^2 \frac{\hat{k}_1}{ \kappa}   } \mathrm{tan}^{-1}\left( \frac{\pi}{2r} \frac{1}{\sqrt{\hat{k}_1^2 - \kappa^2 \mathrm{sin}^2 \frac{\hat{k}_1}{\kappa} }}  \right)  \mathrm{sin}\left[  (2 i'_1+1) \hat{k}_1   \right] 
\end{align}
Numerical results for the analytical expressions are shown in the left column of Fig. \ref{fig: spe}. The $E_1$ field is anti-symmetric and decreases rapidly as one moves away from the particle because of the space charge like nature of the fields. The field structure is insensitive to the particle shape as can be seen by the fact that the quadratic particle shape fields are only slightly less than those for  linear shapes.

While there are no Cerenkov fields from the fundamental zone for a spectral solver, we show such fields exist in the  $\nu_1=\pm1$ zones. We find that the field from the first aliasing zone always has errors from a combination of the Cerenkov and space charge sources. Again we focus on the parameter space where $r=1$ and $\kappa>\kappa_r$ where $\kappa_r\approx 2.4$ can be solved from $\kappa_r^2\mathrm{sin}^2\left( \frac{3}{2}\frac{\pi}{\kappa}\right) = \frac{\pi^2}{2} $. Under this condition, the integration for $\nu_1=1$ can be divided into three regions:
\begin{align}
&-k_{g1}/2<  k_1 \leq -k_{1,r1}: [ k_1 + k_{g1}]^2_t  - k_1^2 \leq 0, \nonumber \\
& \int_{-\frac{k_{g2}}{2}}^{\frac{k_{g2}}{2}} \mathrm{d}k_2  \frac{ [ k_1+k_{g1}]^2_t - k_1^2 } {  [ k_1 + k_{g1}]^2_t  - k_1^2 - k_2^2 } = 2\sqrt{k_1^2 - [k_1+k_{g1}]_t^2} \mathrm{tan}^{-1}\left( \frac{k_{g2}}{2\sqrt{k_1^2 - [k_1+k_{g1}]_t^2}}\right)	\nonumber \\
&-k_{1,r1}<  k_1 \leq -k_{1,r2} : 0 <   [ k_1 + k_{g1}]^2_t  - k_1^2  \leq \frac{k_{g2}^2}{4},  \nonumber \\
& \int_{-\frac{k_{g2}}{2}}^{\frac{k_{g2}}{2}} \mathrm{d}k_2  \frac{ [ k_1+k_{g1}]^2_t - k_1^2 } {  [ k_1 + k_{g1}]^2_t  - k_1^2 - k_2^2 } = 2 \sqrt{ [k_1+k_{g1}]_t^2 - k_1^2 } \left[ -i \frac{\pi}{2} + \mathrm{tanh}^{-1}\left( \frac{2\sqrt{ [k_1+k_{g1}]_t^2 - k_1^2 }}{ k_{g2}} \right)	 \right] \nonumber \\
  &k_1 > -k_{1,r2} :  [ k_1 + k_{g1}]^2_t  - k_1^2  >  \frac{k_{g2}^2}{4}, \nonumber \\
 & \int_{-\frac{k_{g2}}{2}}^{\frac{k_{g2}}{2}} \mathrm{d}k_2  \frac{ [ k_1+k_{g1}]^2_t - k_1^2 } {  [ k_1 + k_{g1}]^2_t  - k_1^2 - k_2^2 } = 2\sqrt{ [k_1+k_{g1}]_t^2 - k_1^2 } \mathrm{tanh}^{-1}\left( \frac{k_{g2}}{2\sqrt{ [k_1+k_{g1}]_t^2 - k_1^2 }}\right)	
\end{align}
where $k_{1,r1}$ and $k_{1, r2}$ satisfy $[ -k_{1,r1} + k_{g1}]^2_t  - k_{1,r1}^2  = 0,  [ -k_{1,r2} + k_{g1}]^2_t  - k_{1,r2}^2 = \frac{k_{g2}^2}{4}$. 

The contribution from the $\nu_1=-1$ term can be calculated similarly. The total  field $E_1$ from the first aliasing modes $\nu_1=\pm1$ can  be shown to be
\begin{align}
&E^n_{1,i_1, i_2=0} (\nu_1=-1) + E^n_{1,i_1, i_2=0} (\nu_1=1) \nonumber \\
&= \frac{q}{\pi^2} \biggl(   \int_{-\frac{k_{g1}}{2}}^{-k_{1,r1}} \mathrm{d}k_1   \frac{ S(k_{1} + k_{g1} )}{ k_1}  \sqrt{ k_1^2 -  [k_1+k_{g1}]_t^2 } \mathrm{tan}^{-1}\left( \frac{k_{g2}}{2\sqrt{k_1^2 - [k_1 + k_{g1}]_t^2 } }\right) \nonumber \\
& \mathrm{sin}\left[ k_1 \left( i'_1+ \frac{1}{2} \right) dx_1 \right]  + \int_{-k_{1,r1}}^{-k_{1,r2}} \mathrm{d}k_1   \frac{ S(k_{1} + k_{g1})}{ k_1}  \sqrt{ [k_1 + k_{g1}]_t^2 - k_1^2   }  \biggl(  \mathrm{tanh}^{-1}\left( \frac{2\sqrt{   [k_1 + k_{g1}]_t^2  - k_1^2 } }{k_{g2}} \right) \nonumber \\ 
&\mathrm{sin}\left[ k_1 \left( i'_1+ \frac{1}{2} \right) dx_1 \right]  - \frac{\pi}{2}  \mathrm{cos}\left[ k_1 \left( i'_1+ \frac{1}{2} \right) dx_1 \right]   \biggr)  + \int_{-k_{1,r2}}^{\frac{k_{g1}}{2}} \mathrm{d}k_1   \frac{  S(k_{1} + k_{g1} )}{ k_1}  \sqrt{ [k_1 + k_{g1}]_t^2 - k_1^2  } \nonumber \\
&\mathrm{tanh}^{-1}\left( \frac{k_{g2}}{2\sqrt{   [k_1 + k_{g1}]_t^2  - k_1^2 } }\right)  \mathrm{sin}\left[ k_1 \left( i'_1+ \frac{1}{2} \right) dx_1 \right] 	\biggr)  \nonumber \\
&= \frac{ 2q}{\pi^2dx_1} \biggl[   \int_{-\frac{\pi}{2}}^{-\hat{k}_{1,r1}} \mathrm{d}\hat{k}_1   \frac{ \hat{S}(\hat{k}_{1} + \pi)}{ \hat{k}_1}  \sqrt{ \hat{k}_1^2 -  [\hat{k}_1+\pi]_t^2 } \mathrm{tan}^{-1}\left( \frac{\pi}{2r\sqrt{\hat{k}_1^2 - [\hat{k}_1 + \pi ]_t^2 } }\right)  \mathrm{sin}[  (2 i'_1+1) \hat{k}_1 ]  \nonumber \\
& +     \int_{-\hat{k}_{1,r1}} ^ {-\hat{k}_{1,r2}} \mathrm{d}\hat{k}_1   \frac{ \hat{S}(\hat{k}_{1} + \pi)}{ \hat{k}_1}  \sqrt{ [\hat{k}_1 + \pi]_t^2 - \hat{k}_1^2   } \biggl( - \frac{\pi}{2} \mathrm{cos}[ (2 i'_1+1) \hat{k}_1 ] +  \mathrm{tanh}^{-1}\left( \frac{2r \sqrt{   [\hat{k}_1 +\pi]_t^2  - \hat{k}_1^2 } } {\pi}\right) \nonumber \\
& \mathrm{sin}[ (2i_1' +1 )\hat{k}_1] \biggr) + \int_{- \hat{k}_{1,r2}}^{\frac{\pi}{2}} \mathrm{d}\hat{k}_1   \frac{  \hat{S}(\hat{k}_{1}+\pi)}{ \hat{k}_1}  \sqrt{ [\hat{k}_1 + \pi]_t^2 - \hat{k}_1^2  } \mathrm{tanh}^{-1}\left( \frac{  \pi }{2r \sqrt{   [\hat{k}_1 +\pi]_t^2  - \hat{k}_1^2 } }\right)  \mathrm{sin}( 2  i'_1 \hat{k}_1)	\biggr]  \nonumber \\
\end{align}
The numerical results are shown in the right column of Fig. \ref{fig: spe}. The Cerenkov radiation pattern at the high wave number region leads to a long tail of the $E_1$ field behind the particle.The use of higher-order particle shapes can significantly reduce the field amplitude by reducing the contributions from the aliasing zones. 

\begin{figure}[htbp]
\begin{center}
\includegraphics[width=0.5\textwidth]{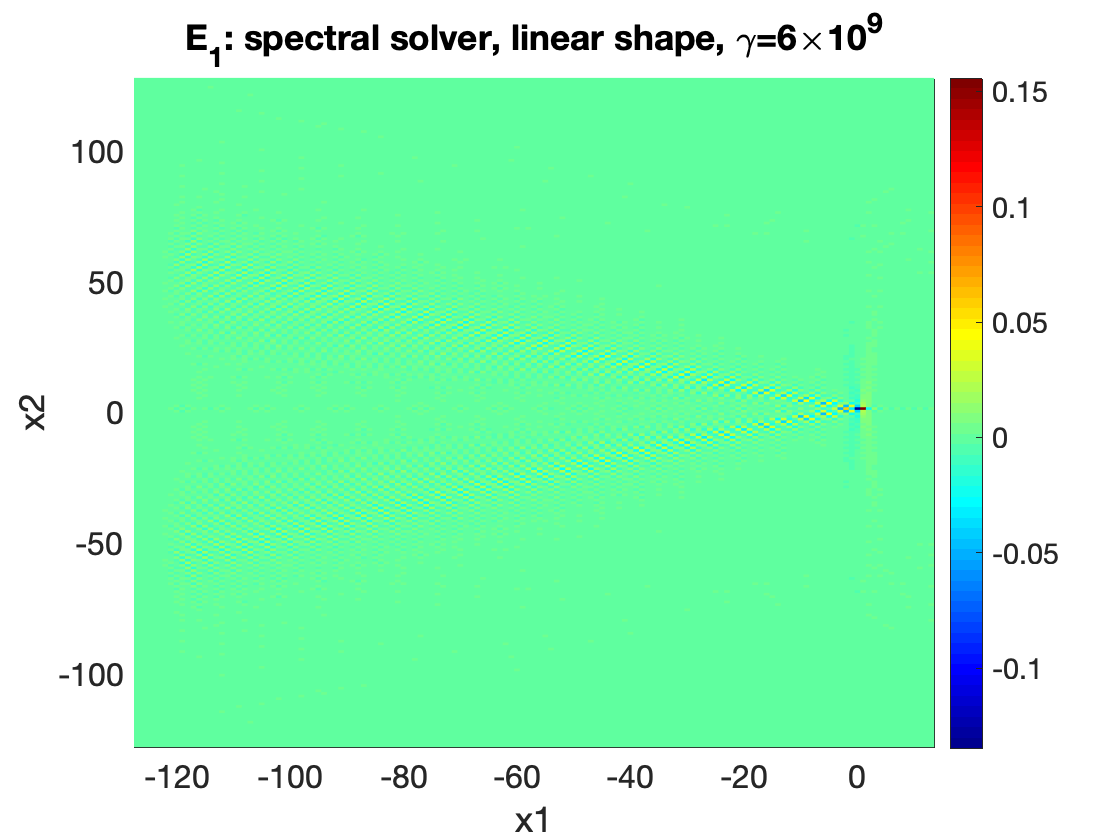}\includegraphics[width=0.5\textwidth]{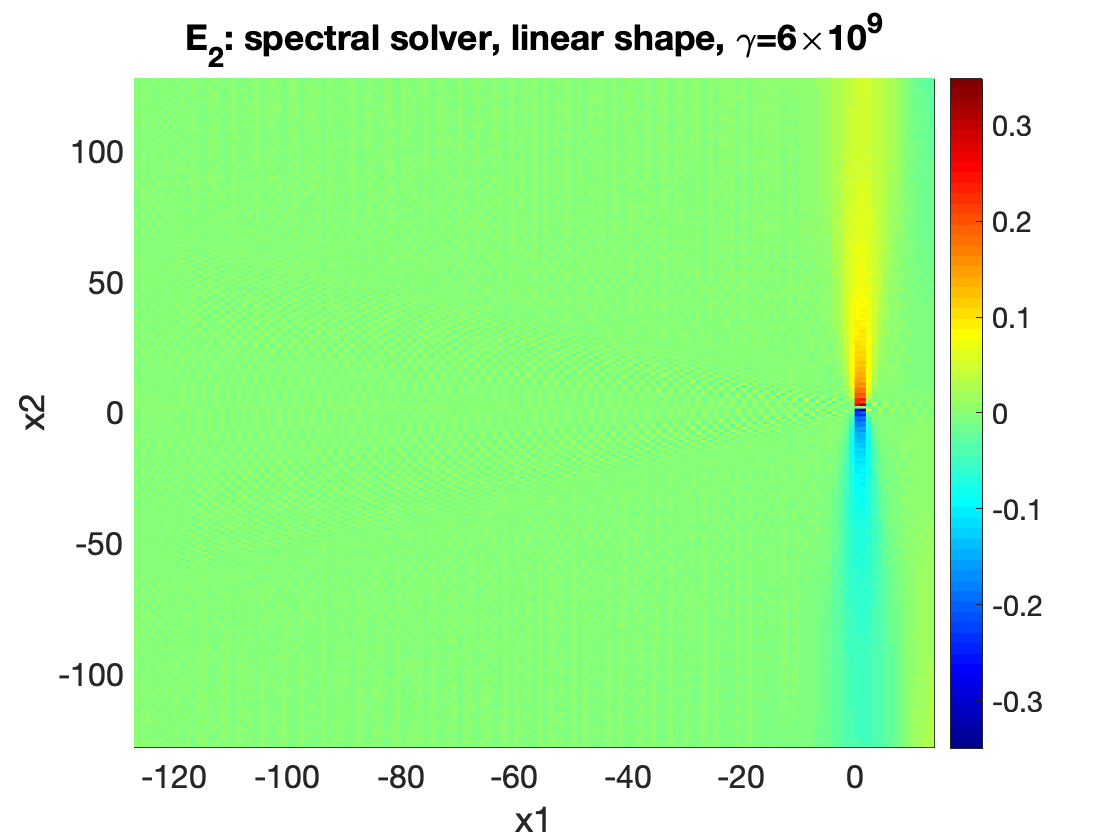}
\includegraphics[width=0.5\textwidth]{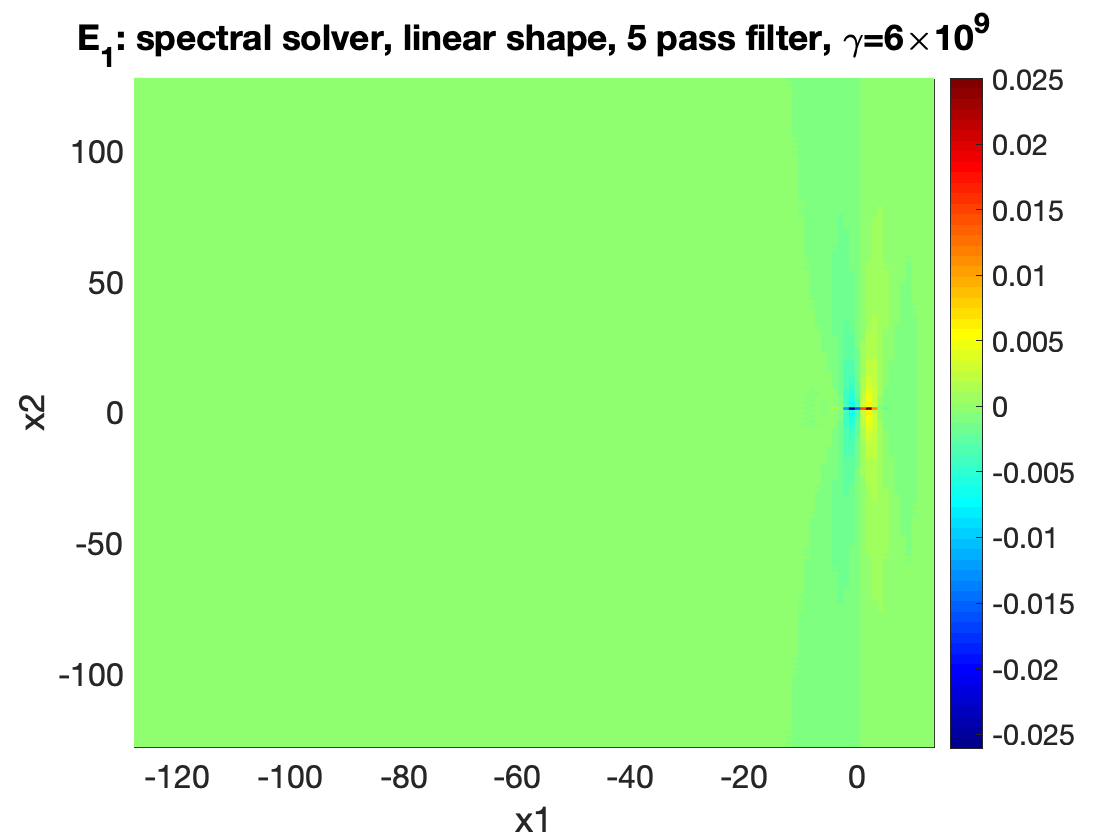}\includegraphics[width=0.5\textwidth]{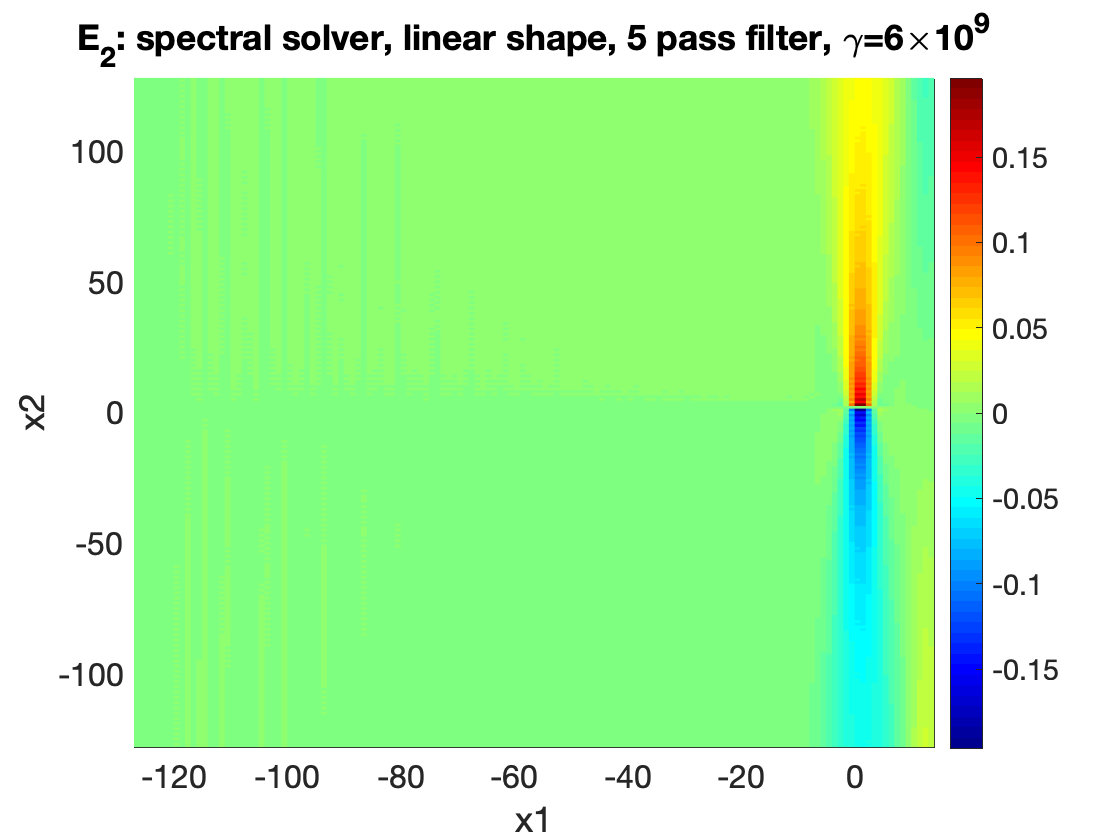}
\includegraphics[width=0.5\textwidth]{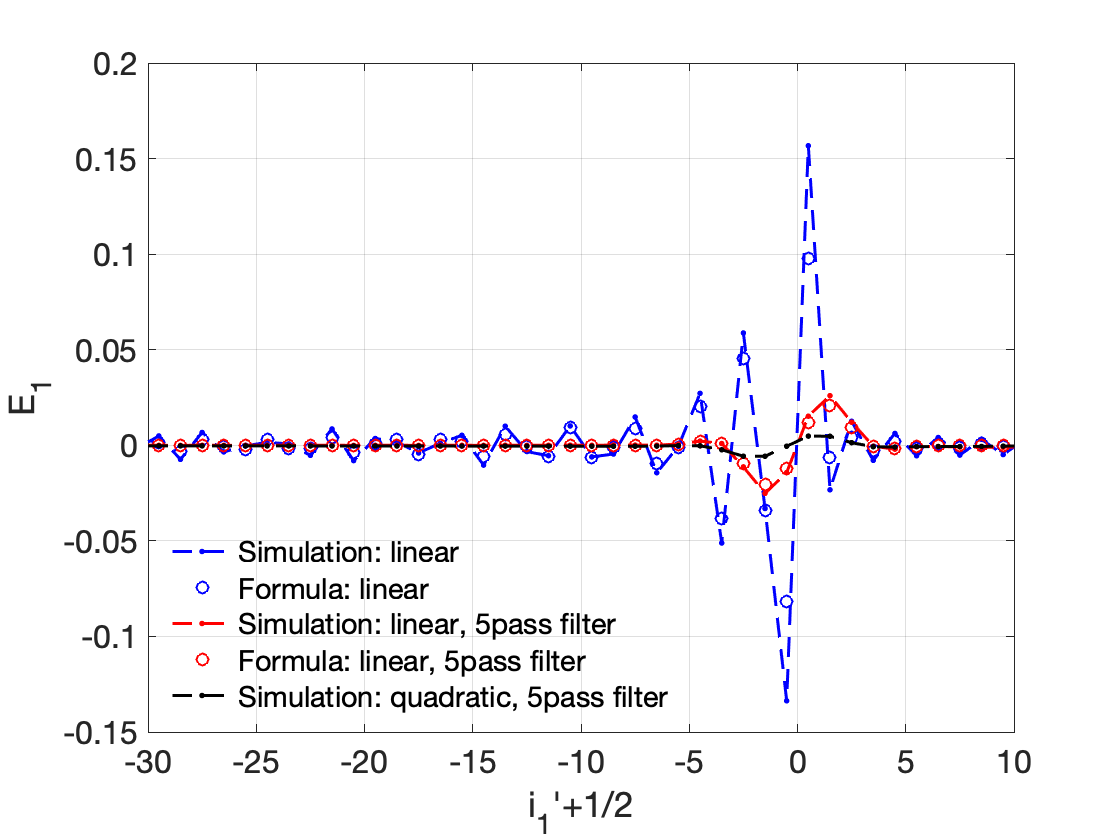}\includegraphics[width=0.5\textwidth]{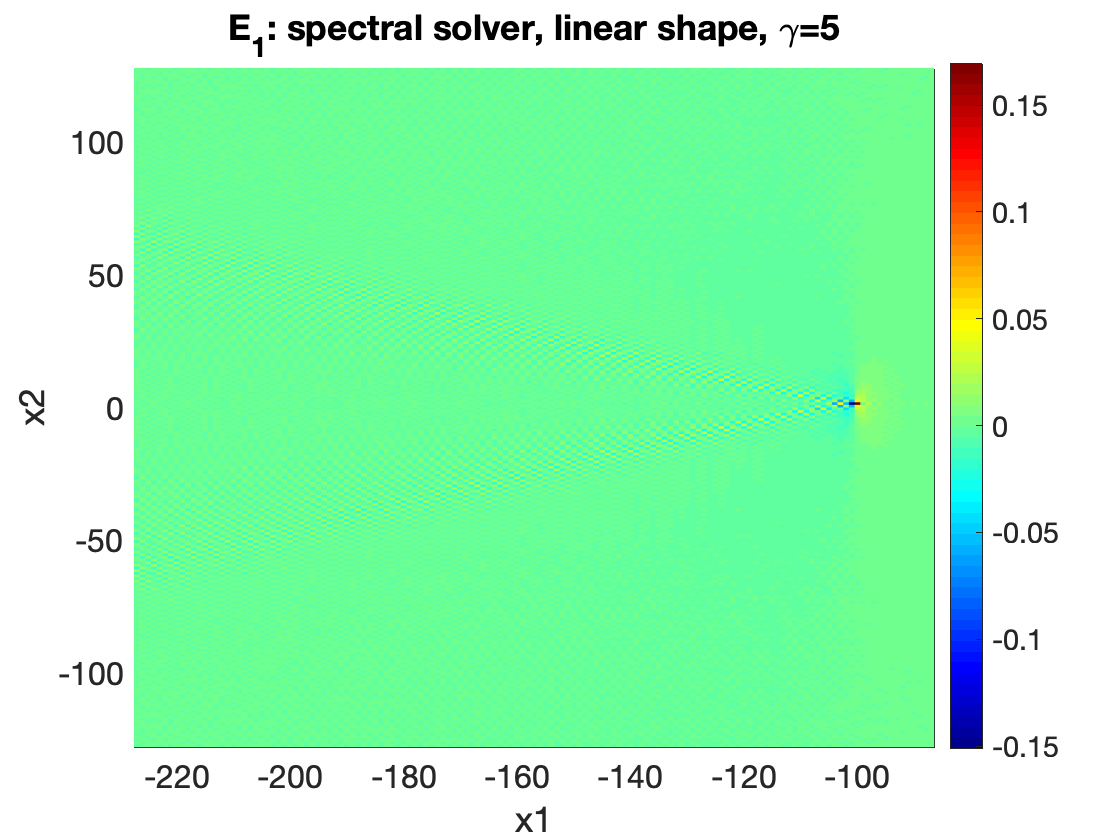}
\caption{The electrical fields $E_1$ and $E_2$ of a free-streaming particle from PIC simulations with the spectral solver. Parameters: $dx_1=1, r\equiv \frac{dx_2}{dx_1}=1, \kappa\equiv \frac{dx_1}{dt}=4, q=1$. Other simulation parameters can be found in each subplots.}
\label{fig: spe sim}
\end{center}
\end{figure}

OSIRIS simulation results are shown in Fig. \ref{fig: spe sim}. Compared with the Yee solver, the fields are smaller and are now dominated by the numerical space charge mode, thus the numerical fields have significant values only close to the particle. In addition, the numerical Cerenkov mode exist only in the high $k_1$ region and thus a 5 pass filter can reduce it significantly. Reasonable agreement between the simulation results and the analytic expressions can be seen in the bottom left of Fig. 4 where the disparity near the particle position is likely due to the neglect of the contributions from higher $\nu_1$ ($\nu_1 \geq 2$) zones. Combining quadratic particle shapes with a 5 pass filter, the $E_1$ field ( black dashed line) is on the order of $E_1 \frac{dx_1}{q} \sim 5\times10^{-3}$. The $E_1$ field distribution from a particle with $\gamma=5$  (relatively low energy) is shown in the bottom right of Fig. 4. We can see it is similar to the field from a very relativistic particle as shown in the top left of Fig. 4 which confirms that  even with a spectral solver the self-fields are also dominated by the numerical effects. 

The self-force experienced by a particle with charge $q$ for an spectral solver is 
\begin{align}
F_1 \approx -\frac{q^2}{\pi dx_1 } \int_{-\hat{k}_{1,r1}}^{-\hat{k}_{1,r2}} \mathrm{d}\hat{k}_1 \frac{\hat{S}(\hat{k}_1)\hat{S}(\hat{k}_1 + \pi)}{\hat{k}_1} \sqrt{ [\hat{k}_1 + \pi]_t^2 - \hat{k}_1^2   } 
\end{align}
where only the contributions from $\nu_1=0, \pm1$ are considered. The value of the force can be integrated numerically leading to $F_1\approx 0.016 q^2/dx_1$ for linear particle shape, $F_1\approx 0.0063 q^2/dx_1$ for quadratic shape and $F_1\approx 6.5\times10^{-6}q^2/dx_1$ for quadratic shape and a 5 pass filter when $ r=1, \kappa=4$.

We close this section by noting that it is more difficult to carry out a similar analysis for the PSATD approach. It is difficult to cast the solver into a simple form where $[\omega]$ and $[k]$'s can be used. In \ref{sec: psatd} we provide an analysis which provides expressions for the Fourier amplitudes of the electric field and magnetic fields in terms of $\bm {k}$, $k dt$ and $\omega dt$. An expression for axial electric field is then provided which has similar poles in the denominator as in the spectral solver. However, it is difficult to carry out the integral in Eq. \ref{eq: B5}.

\section{Solution: a solver with $[k]_1 = [k_1]_t$ }
\label{sect:solution}
As explained above, the unphysical fields are mainly caused by the different forms of $[.]_t$ and $[.]_1$ for the solvers. Thus, we propose a new solver with $[.]_1 = [.]_t$ which can significantly reduce the numerical self fields of relativistic particles below those from the spectral solver. For simplicity, we assume $[k]_{E1} = [k]_{B1}$, other options with $[k]_{E1}\neq [k]_{B1}$ and $[k]_{E1}[k]_{B1}=[k_1]_t^2$ are possible. From Eq. \ref{eq: EM} it is straightforward to see that using $[k]_{E1} = [k]_{B1}$ has another advantage in that the  transverse force between two relativistic particles is also free of additional numerical errors, i.e.,$ E_2-\beta B_3$=$E_2 (1-\beta)$. This will be discussed in more detail in a future publication. 

To see the advantage of the new solver, we do the same analysis for the new solver as in Eq. (\ref{eq: taylor expansion}),
\begin{align}
\frac{[\beta k_1]_t^2 - [k]_1^2 }{k_1^2 } = -\frac{1}{\gamma^2}\left[  1 - \frac{k_1^2 dt^2}{6} + O((k_1dt)^4) \right]  + O\left(\frac{1}{\gamma^4}\right)
\end{align} 
When $k_1^2 \frac{ dt^2}{  6} \ll 1 $, i.e., $|k_1 \frac{dx_1}{2}| \ll \frac{\sqrt{6}}{2}\frac{dx_1}{dt}$, the physical fields are modeled well on the grids. Compared with Eq. (\ref{eq: Taylor expansion}), the range of $k_1$ where the fields are modeled with high fidelity  is much increased. The contributions from  aliasing (higher order Brillouin zones) to the fields still exist, however they are concentrated at the high $k_1$ region which can be suppressed using high-order particle shapes and low pass filters. 

The contribution to the $E_1$ field from the first aliasing modes $\nu_1=\pm1$ can be calculated as follows. When $\nu_1=\pm1$ the fields can have different characteristics depending on the value of $\kappa$ and $r$. Here we focus on $r=1$ and $\kappa\geq \kappa_r$ where $\mathrm{sin}^2(\frac{3\pi}{2\kappa_r}) - \mathrm{sin}^2(\frac{\pi}{2\kappa_r}) \equiv 1/\kappa_r^2$ with $\kappa_r\approx 2.15$. Taking $\nu=1$ as an example, the integration can be done as follows:
\begin{align}
&-k_{g1}/2 < k_1 \leq -k_{1,r}: 0 < [k_1+k_{g1}]_t^2 - [k_1]_t^2 \leq \left( \frac{dx_2}{2} \right)^{-2}, \nonumber \\
&\int_{-k_{g2}/2}^{k_{g2}/2} \mathrm{d} k_2 \frac{[k_1 + k_{g1}]_t^2 - [k_1]_t^2}{[k_1 + k_{g1}]_t^2 - [k_1]_t^2 - [k]_2^2} = -i\pi \sqrt{ \frac{[k_1+k_{g1}]_t^2 - [k_1]_t^2}{1-\left( \frac{dx_2}{2}\right)^2 ([k_1+k_{g1}]_t^2 - [k_1]_t^2)}}  \nonumber \\
& -k_{1,r} < k_1 \leq k_{g1}/2: [k_1+k_{g1}]_t^2 - [k_1]_t^2 >  \left( \frac{dx_2}{2} \right)^{-2},  \nonumber \\
&\int_{-k_{g2}/2}^{k_{g2}/2} \mathrm{d} k_2 \frac{[k_1 + k_{g1}]_t^2 - [k_1]_t^2}{[k_1 + k_{g1}]_t^2 - [k_1]_t^2 - [k]_2^2} =\pi \sqrt{ - \frac{[k_1+k_{g1}]_t^2 - [k]_1^2}{1-\left( \frac{dx_2}{2}\right)^2 ([k_1+k_{g1}]_t^2 - [k_1]_t^2)}}
\end{align}
where $[-k_{1,r} + k_{g1}]_t^2 - [k_{1,r}]_t^2=\left( \frac{dx_2}{2} \right)^{-2}$. Then the field is
\begin{align}
&E_{1,i_1,i_2=0}^n(\nu_1=-1) + E_{1,i_1,i_2=0}^n(\nu_1=1) \nonumber \\
&= -\frac{q}{2\pi} \biggl[ \int_{- \frac{k_{g1}}{2}}^{-k_{1,r}} \mathrm{d} k_1 \frac{ S_{1}(k_1 + k_{g1}) }{[k_1]_t} \sqrt{\frac{[k_1+k_{g1}]_t^2 - [k]_t^2}{1-\left(\frac{ dx_2}{2}\right)^2 ([k_1 + k_{g1}]_t^2 - [k_1]_t^2)}} \mathrm{cos} \left[k_1dx_1 \left(i_1'+\frac{1}{2}\right)\right]\nonumber \\
&+ \int_{-k_{1,r}}^\frac{k_{g1}}{2}\mathrm{d}k_1 \frac{S_1(k_1 + k_{g1}) }{[k_1]_t}\sqrt{ -\frac{[k_1+k_{g1}]_t^2 - [k_1]_t^2}{1-\left( \frac{dx_2}{2}\right)^2 ([k_1+k_{g1}]_t^2 - [k_1]_t^2)}} \mathrm{sin}\left[k_1dx_1 \left(i_1'+\frac{1}{2} \right) \right] \biggr]	\nonumber \\
=&-\frac{q }{\pi dx_1} \biggl[ \int_{- \frac{\pi}{2}}^{-\hat{k}_{1,r}} \mathrm{d} \hat{k}_1 \frac{\hat{S}_{1}( \hat{k}_1 + \pi )}{ \mathrm{sin}\frac{\hat{k}_1}{\kappa} } \sqrt{\frac{ \kappa^2 \left( \mathrm{sin}^2\frac{\hat{k}_1 + \pi}{\kappa}  - \mathrm{sin}^2\frac{\hat{k}_1}{\kappa} \right)  }{1-r^2\kappa^2 \left( \mathrm{sin}^2\frac{\hat{k}_1 + \pi}{\kappa}  - \mathrm{sin}^2\frac{\hat{k}_1}{\kappa} \right)}}  \mathrm{cos}[\hat{k}_1(2i_1'+1)]\nonumber \\
&+ \int_{- \hat{k}_{1,r}}^\frac{\pi}{2}\mathrm{d} \hat{k}_1 \frac{\hat{S}_{1}( \hat{k}_1 + \pi )}{ \mathrm{sin}\frac{\hat{k}_1}{\kappa} } \sqrt{ - \frac{ \kappa^2 \left( \mathrm{sin}^2\frac{\hat{k}_1 + \pi}{\kappa}  - \mathrm{sin}^2\frac{\hat{k}_1}{\kappa} \right)  }{1-r^2\kappa^2 \left( \mathrm{sin}^2\frac{\hat{k}_1 + \pi}{\kappa}  - \mathrm{sin}^2\frac{\hat{k}_1}{\kappa} \right)}} \mathrm{sin}[\hat{k}_1(2i_1'+1)]\biggr]
\end{align}
where  staggered grids are used. We can see the fields are dominated by the space charge pattern. Higher-order particle shapes can suppress the numerical errors that arise from the aliasing zones. The self-forces can be calculated numerically as $F_1\approx 0.040 q^2/dx_1$ for linear particle shape, $F_1\approx 0.016 q^2/dx_1$ for quadratic shape and $F_1\approx 3.3\times 10^{-8}q^2/dx_1$ for quadratic shape and a 5 pass filter when $r=1, \kappa=4$.

\begin{figure}[htbp]
\begin{center}
\includegraphics[width=0.5\textwidth]{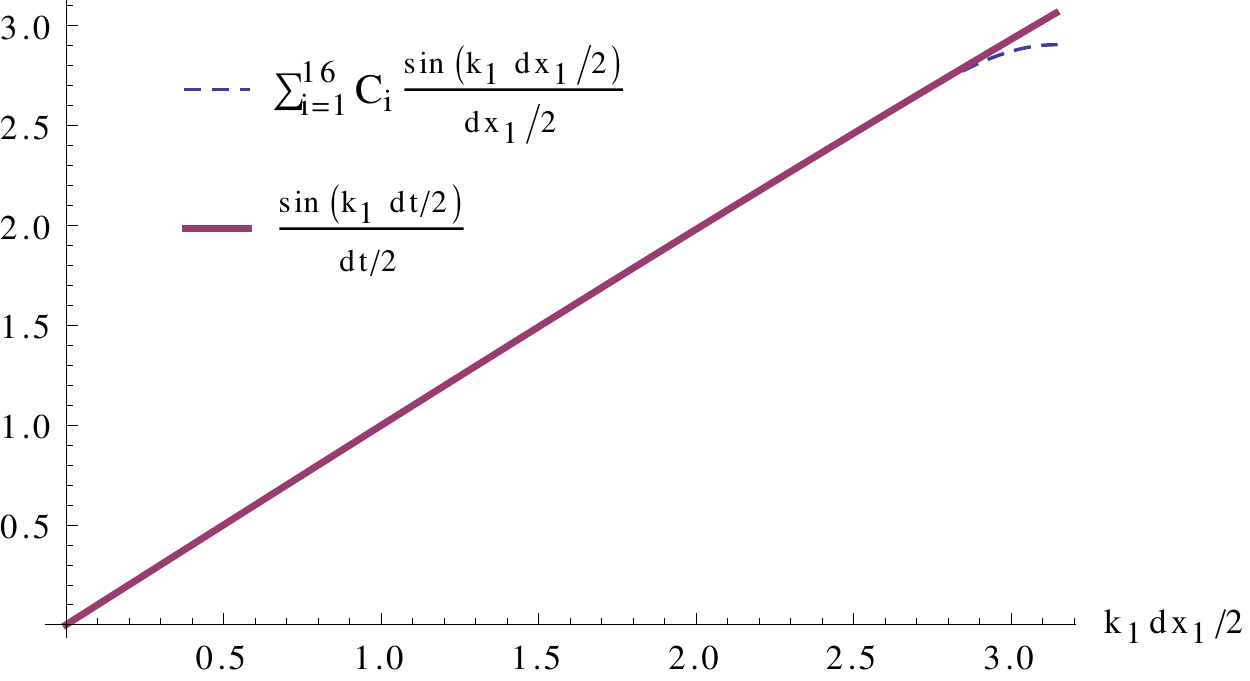}\includegraphics[width=0.5\textwidth]{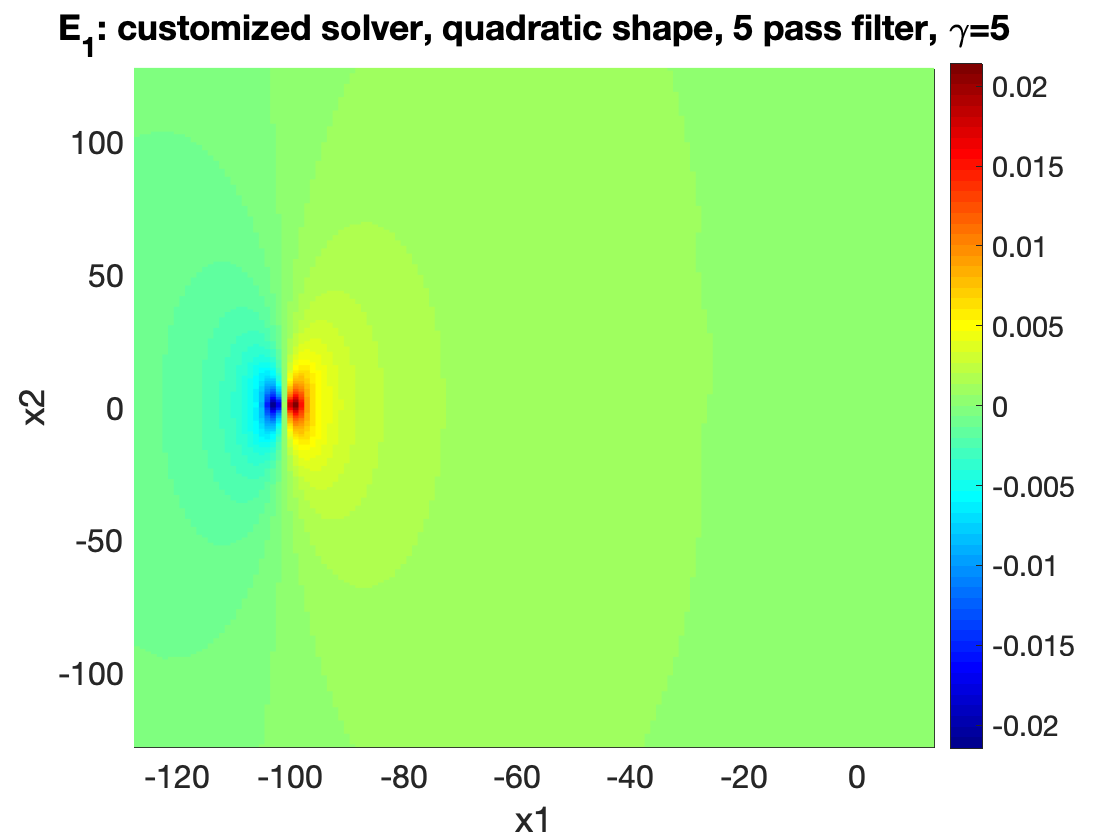}
\includegraphics[width=0.5\textwidth]{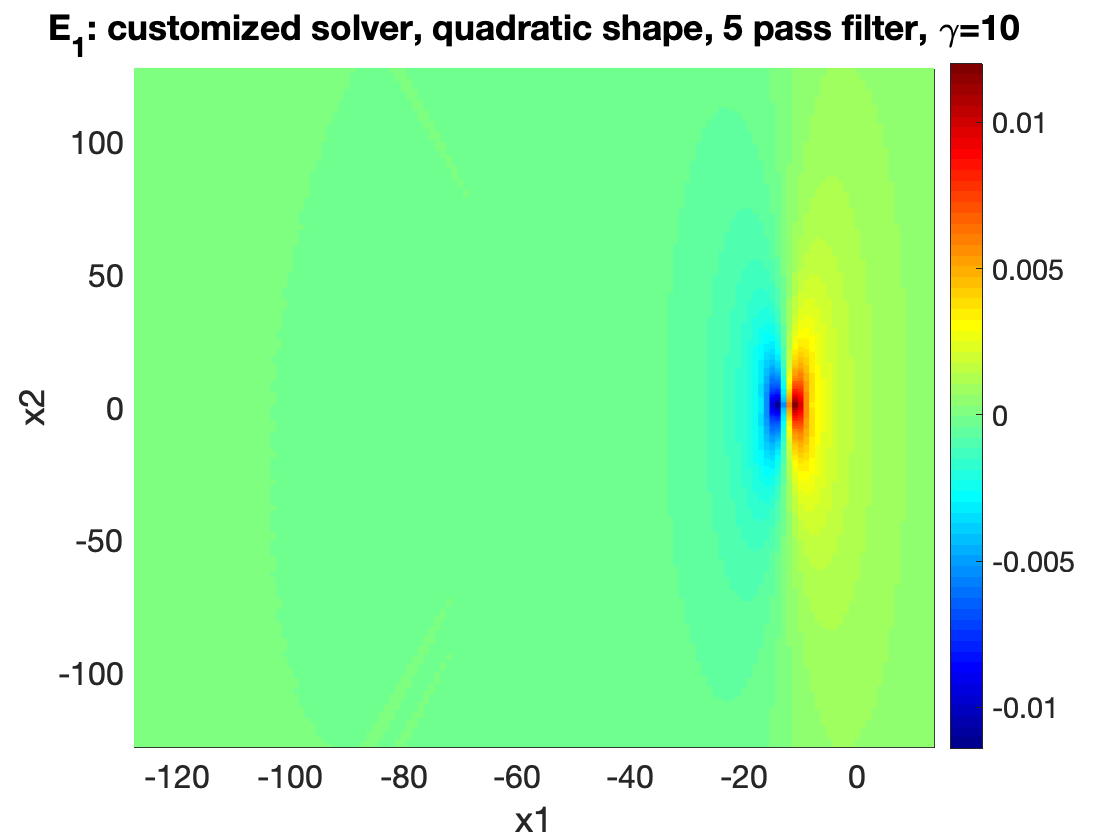}\includegraphics[width=0.5\textwidth]{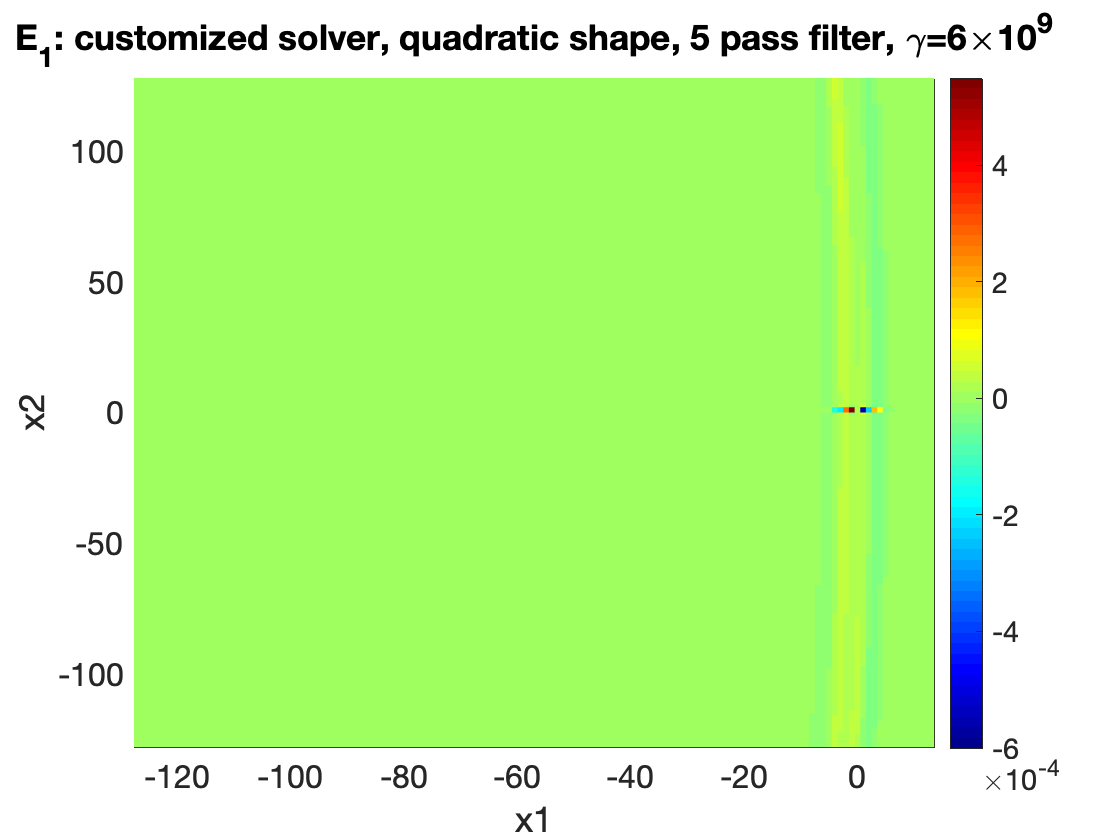}
\includegraphics[width=0.5\textwidth]{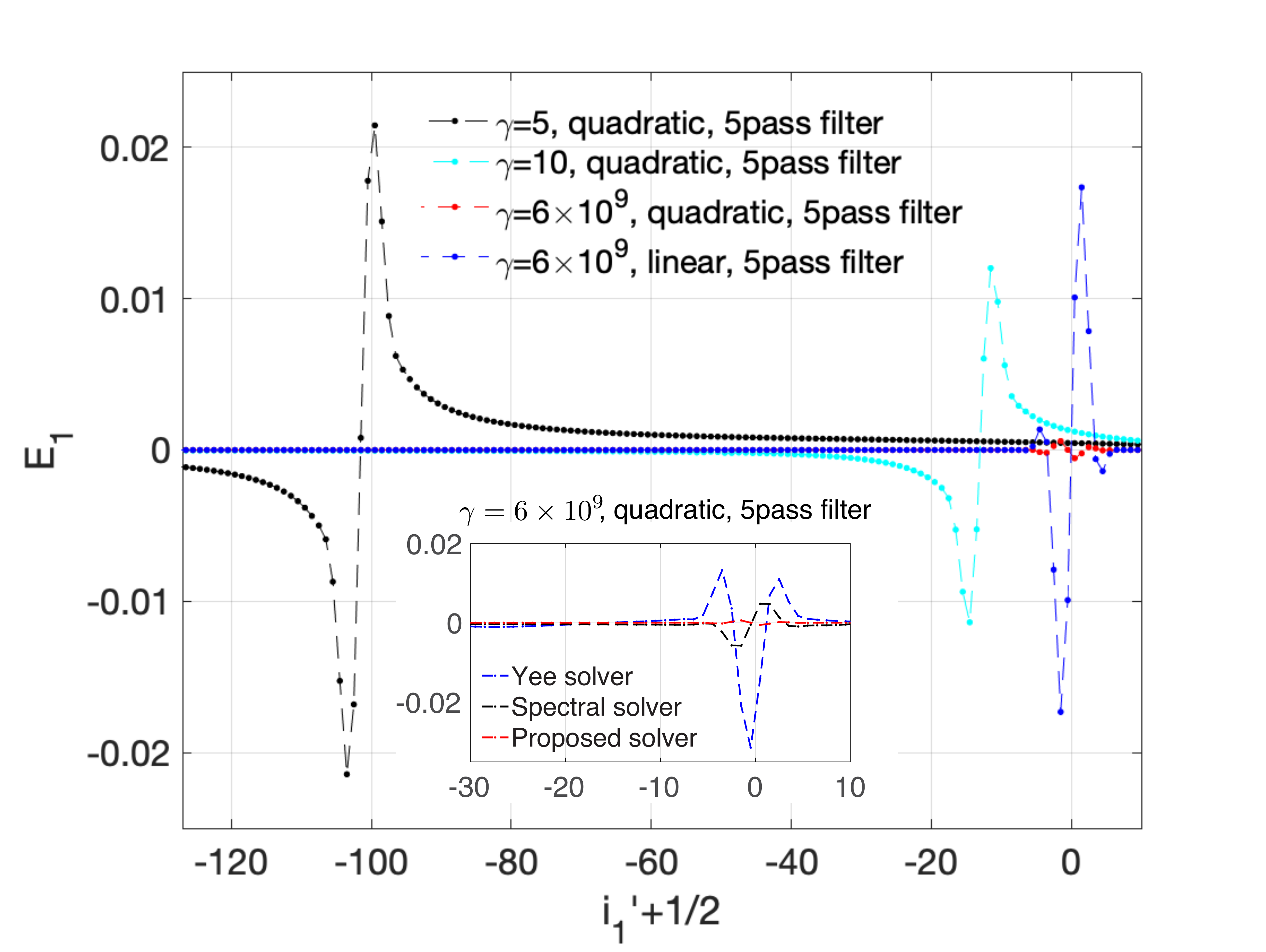}\includegraphics[width=0.5\textwidth]{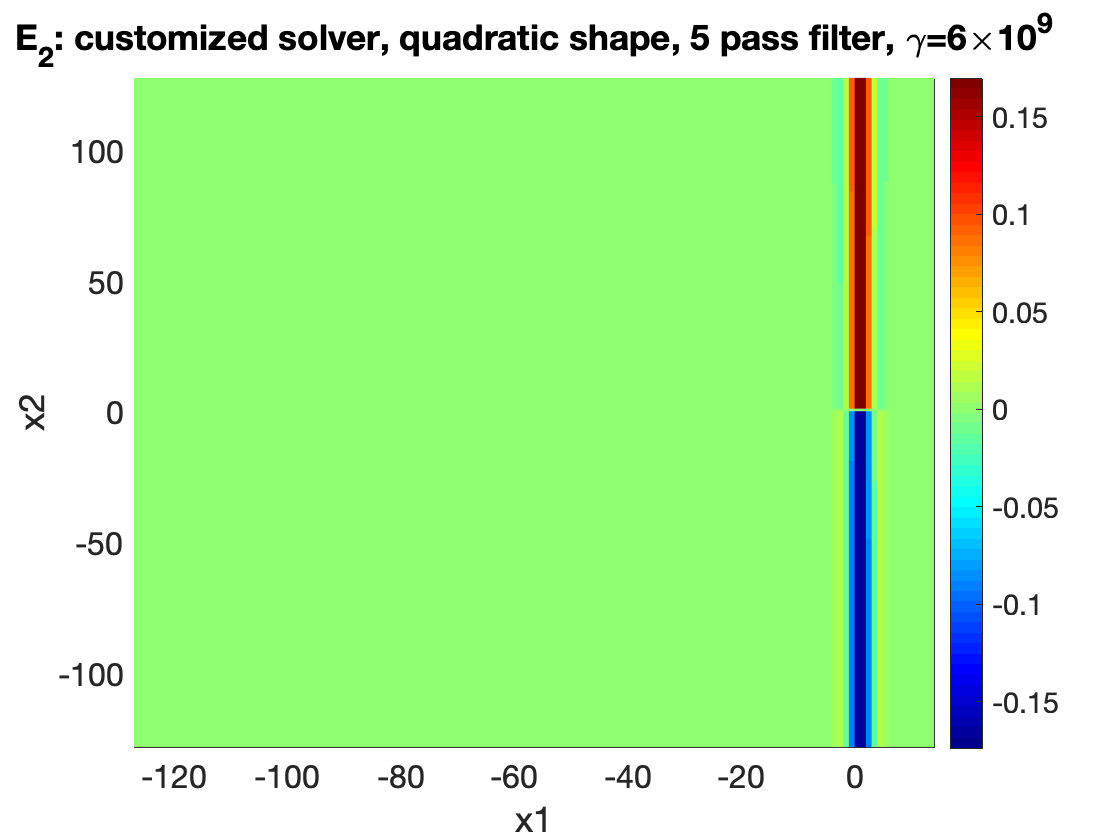}
\caption{Upper left: The $[k]_1$ from the customized solver with the coefficients given in Table. 1 and its comparison with $\mathrm{sin}(k_1dt/2)/(dt/2)$; $E_1$ field when $\gamma=5$ (upper right), $\gamma=10$ (middle left) and $\gamma=6\times10^9$ (middle right); Bottom left: the on-axis lineout of $E_1$ for different $\gamma$ and particle shapes and the inset compares the results from different solvers for the same $\gamma$, particle shape and filter; Bottom right: $E_2$ field when $\gamma=6\times10^9$. Parameters: $dx_1=1, r\equiv \frac{dx_2}{dx_1}=1, \kappa\equiv \frac{dx_1}{dt}=4, q=1$. Other parameters are shown in each subplot.}
\label{fig: cus sim}
\end{center}
\end{figure}

The proposed solver with $[.]_1=[.]_t$ can be easily implemented if the fields are advanced in $\bm{k}$ space. For finite difference solvers, one can use the customized finite difference solver \cite{li2017controlling} technique to approximate $[.]_1=[.]_t$. By extending the stencil of the finite difference operator from 2 grids as in the Yee solver to $2M$ grids, i.e., from $\mathrm{d}_{x1} f_{i_1} = \frac{f_{i_1+1}-f_{i_1}}{dx_1}$ to $\mathrm{d}_{x1} f_{i_1}= \sum_{l=1}^M C_l \frac{f_{i_1+l} - f_{i_1-l+1}}{dx_1}$, the customized solver has a $k_1$ space operator $[k]_1 = \sum_{l=1}^{M} C_l \frac{\mathrm{sin}[(2l-1)k_1 dx_1/2]}{dx_1/2}$. The coefficients $C_l$ are chosen so that $[k_1]$ has accuracy to a chosen order and to minimize errors from the desired functional form for $[k_1]$. Here we take $\kappa=4, r=1$ as an example and $M=16$, the optimized coefficients are listed in Table. \ref{tab:coefs} and the corresponding $[k]_1$ is shown in the upper left of Fig. \ref{fig: cus sim}. We can see  $[k]_1$ is very close to $[k_1]_t$ except at the very high $k_1$ region. The numerical EM fields from this high $k_1$ region can be suppressed with a low pass filter. The scheme to find the customized coefficients that minimize errors to the desired operator can be found in Appendix C. The current is corrected corresponding to the customized coefficients $C_l$ to ensure the Gauss's law $\bm{d} \cdot \bm{E} = \rho$  as described in Ref. \cite{li2017controlling}.   

The simulation results from OSIRIS are shown in Fig. \ref{fig: cus sim}. The fields are modeled well for particles with energy $\gamma=5, 10 $ and $6\times10^9$. The numerical $E_1$ field from the first aliasing zones can be seen when $\gamma=6\times10^9$ and its amplitude is $\sim6\times 10^{-4}$ while the physical field is close to zero. A comparison of the on-axis $E_1$ field for different particle energies and particle shapes is shown in the bottom left of Fig. \ref{fig: cus sim}. Quadratic particle shapes reduce the numerical field by an order of amplitude as compared with the linear shape. We also compare the on-axis $E_1$ field for a particle with $\gamma=6\times 10^9$, the quadratic particle shape and a 5 pass filter using different solvers in the inset. The $E_2$ field for an ultra-relativistic particle is shown in the bottom right which is also modeled well. 

A summary of the types of the numerical errors surrounding a relativistically moving particle for different solvers in the fundamental and first aliasing zones is given in Table \ref{tab:summary}. The value of the self-forces is also shown.

\begin{table}[t]
\centering
\begin{tabular}{cccc}
\hline\hline
\textbf{Coefficients} & \textbf{Values} & \textbf{Coefficients} & \textbf{Values} \\ 
\hline
$C_1$ & 1.248130933469396 & $C_2$ & -0.125139446419605 \\
$C_3$ &  0.040951412341664 & $C_4$ & -0.018181000724779 \\
$C_5$ & 0.009121734802653 & $C_6$ &  -0.004812019139855 \\
$C_7$ & 0.002570175265005 & $C_8$ & -0.001356670416082 \\
$C_9$ &   0.000694723529383 & $C_{10}$ &  -0.000339356036052 \\
$C_{11}$ &  0.000155316438379 & $C_{12}$ & -0.000065152281459 \\
$C_{13}$ & 0.000024294488165 & $C_{14}$ &  -0.000007666519708  \\
$C_{15}$ & 0.000001867004186 & $C_{16}$ & -0.000000273002422 \\
\hline\hline
\end{tabular}
\caption{Coefficients $C_i$ for the customized solver when $\kappa=4, r=1$. }
\label{tab:coefs}
\end{table}

\subsection{A sample simulation: a relativistic beam drifts in free space}
\begin{figure}[htbp]
\begin{center}
\includegraphics[width=0.5\textwidth]{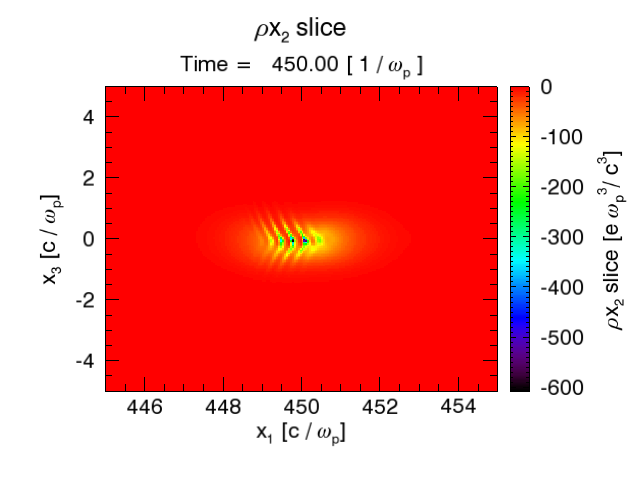}\includegraphics[width=0.5\textwidth]{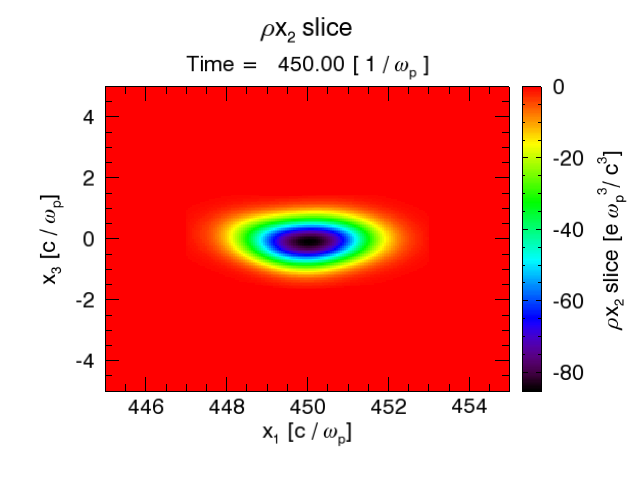}
\caption{Comparison of the beam evolution in free space using Yee solver (left) and our new solver (right). A tri-gaussian beam with $E_b=1~\mathrm{GeV}, n_b=100n_p, k_p\sigma_z=1, k_p\sigma_r=0.5$ and zero emittance and energy spreads propagates in free space. The charge density distribution of beam at $\omega_p t=450$ are shown. Parameters: $dx_1=0.05, r\equiv \frac{dx_2}{dx_1}=1, \kappa\equiv \frac{dx_1}{dt}=4$, and there are 8 particles per cell for representing the beam.}
\label{fig:density plots}
\end{center}
\end{figure}

Here we give a comparison of the evolution of an electron beam when it drifts in free space using the Yee solver and our proposed solver in 3D geomerty. The density plots in $x_2=0$ slice at $\omega_p t=450$ are shown in Fig. \ref{fig:density plots} where the beam with the Yee solver breaks into several beamlets and the beam with our solver remains the same as $t=0$.

\begin{table}[t]
\centering
\begin{tabular}{ccccc}
\hline\hline
\textbf{} & \textbf{Yee} & \textbf{Spectral }  &  \textbf{solver with $[k]_1=[k_1]_t$} \\ 
\hline
\multirow{2}{4em}{ $\nu_1=0 $} & C + SC& SC & None \\
				     & $F_1\approx -0.031 q^2/dx_1$      & $F_1=0$      &  $F_1=0$         \\		
\hline
\multirow{2}{4em}{ $|\nu_1|=1$ } &  SC & C + SC&   C+SC \\
						  &   $F_1=0$     & $F_1 \approx 6.5\times 10^{-6} q^2/dx_1$   & $F_1\approx 3.3\times 10^{-8} q^2/dx_1$\\
\hline\hline
\end{tabular}
\caption{ Types of numerical errors for different solvers in the fundamental and first aliasing zones when $\kappa=4, r=1$, where C and SC represent Cerenkov like fields and space charge like fields, respectively. $F_1$ is the self-force along the particle moving direction where quadratic particle shape and a 5 pass filter are used. }
\label{tab:summary}
\end{table}

\section{Summary}
\label{sect:summary}
Modeling relativistic charged particles with high fidelity in PIC codes is important for both beam and plasma physics. In this article, we analyzed the numerical errors to the fields that surround a relativistically charged particle that  free streams across the grid. Two types of errors are identified, one is from Cerenkov like radiation that arises in the fundamental Brillouin zone when the phase velocity of light for the Maxwell solver is less than the speed of light and that arises in the higher order zones regardless of the solver. The other type is a space charge like field that arises when the errors in the finite difference operators in time and position are larger than $1\/gamma^2$. The details of these errors are analyzed for finite difference and FFT based solvers analytically and in PIC simulations. A novel solver with $[k]_1=[k_1]_t$ is proposed and implemented in OSIRIS to eliminate these numerical fields. The simulation results with the proposed solver show the fields are modeled well and the amplitude of the numerical errors is reduced by one order of magnitude compared with the Yee solver and the spectral solver. 

Areas for future work include developing methods for mitigating these errors when particles are streaming simultaneously at arbitrary angles and to better understand how these errors self-consistently cause coherent interactions within a beam. 

\section*{Acknowledgments}
Work supported by the U.S. Department of Energy under contract number DE-AC02-76SF00515, No. DE-SC0010064, and SciDAC FNAL subcontract 644405, and NSF Grants Nos. 1734315 and ACI-1339893,
Fundação para a Ciência e Tecnologia (FCT), Portugal, grant no. PTDC/FIS-PLA/2940/2014. The simulations were performed on the UCLA Hoffman 2 and Dawson 2 Clusters, and the resources of the National Energy Research Scientific Computing Center.

\begin{appendix}

\section{Free streaming charge density}
\label{sec: AppendixA}
When the particles are free streaming along the $x_1$ direction, the charge density on the grids at the $n$ time step can be written as
\begin{align}
\rho_{i_1,i_2,i_3}^{n} &=\int_{-\infty}^{+\infty} \mathrm{d}^3\bm{x} \rho^{n} (x_1, x_2,x_3)S(i_1 dx_1 - x_1, i_2 dx_2 - x_2, i_3 dx_3 - x_3 ) \nonumber \\
&=\int_{-\infty}^{+\infty}  \mathrm{d}^3\bm{x} \rho^{0} (x_1-n\beta dt, x_2, x_3) S(i_1 dx_1 - x_1, i_2 dx_2 - x_2, i_3 dx_3 - x_3 ) 
\end{align}
where $\rho^0(\bm{x})$ is the charge density of the particles at $t=0$ and $S(\bm{x})$ is the particle shape function. Then apply the Fourier transform in the time domain and the $x_1$ space domain to the above expression,
\begin{align}
\tilde{\rho}(\omega, k_1) &= \sum_{n,i_1} \rho^n_{i_1,i_2,i_3} \mathrm{exp}(-i k_1 i_1 dx_1) \mathrm{exp}\left( i \omega n dt \right) \nonumber \\
&= \sum_{n,i_1} \int_{-\infty}^{+\infty} \mathrm{d}x_1 \rho^0(x_1-n\beta dt) S(i_1 dx_1 - x_1) \mathrm{exp}(-i k_1 i_1 dx_1) \mathrm{exp}\left( i \omega n dt \right) 	\nonumber \\
&= \sum_{n,i_1}  \int_{-\infty}^{+\infty} \mathrm{d}x_1   S(i_1 dx_1 - x_1)    \int_{-\infty}^{+\infty} \frac{\mathrm{d}k'_1}{2\pi}    \tilde{\rho}^0(k'_1) \mathrm{exp} [i k'_1 (x_1 - n\beta dt)]  \mathrm{exp}\left(-i k_1 i_1 dx_1 + i \omega n dt \right) 	\nonumber \\
&=  \int_{-\infty}^{+\infty} \frac{\mathrm{d}k'_1}{2\pi} \tilde{\rho}^0(k'_1)   \sum_{n,i_1} \int_{-\infty}^{+\infty} \mathrm{d}x_1 S(i_1 dx_1 - x_1 )  \mathrm{exp} \left[i k'_1 (x_1 - n\beta dt) -i k_1 i_1 dx_1 + i \omega n dt \right] 	\nonumber \\
&= \int_{-\infty}^{+\infty} \frac{\mathrm{d}k'_1}{2\pi} \tilde{\rho}^0(k'_1) S(k'_1)  \sum_{i_1} \mathrm{exp}(ik'_1i_1dx_1 - i k_1 i_1 dx_1 ) \sum_{n} \mathrm{exp}\left(-ik'_1 n \beta dt + i \omega n dt \right) 	\nonumber \\
&= \int_{-\infty}^{+\infty} \frac{ \mathrm{d}k'_1 }{2\pi}\tilde{\rho}^0(k'_1)  S(k'_1)   \frac{2\pi}{dx_1}\sum_{\nu_1} \delta\left[k'_1 - ( k_1 + \nu_1 k_{g1}) \right]  \frac{2\pi}{dt} \sum_\mu \delta \left( \omega+ \mu \omega_g - \beta k'_1 \right) 	\nonumber \\
&= \frac{2\pi}{dt dx_1 } \sum_{\mu, \nu_1} \tilde{\rho}^0(k_{1} + \nu_1 k_{g1})  S( k_{1} + \nu_1 k_{g1})  \delta \left[ \omega+ \mu \omega_g - \beta (k_1+\nu_1 k_{g1} ) \right]
\end{align}
where $k_{g1}=\frac{2\pi}{dx_1}$ and $\omega_{g}=\frac{2\pi}{dt}$. After applying the Fourier transform along the $x_2$ and $x_3$ directions, the expression of $\tilde{\rho}$ is
\begin{align}
\tilde{\rho}(\omega, \bm{k}) = \frac{2\pi}{dt dx_1 dx_2 dx_3} \sum_{\mu, \bm{\nu}} \tilde{\rho}^0(\bm{k}')  S( \bm{k}')  \delta \left[ \omega+ \mu \omega_g - \beta (k_1+\nu_1 k_{g1} ) \right] \label{eq:rho}
\end{align}
where $k'_{1,2,3} = k_{1,2,3} +\nu_{1,2,3} k_{g1,2,3}$ and $k_{g2,3}=\frac{2\pi}{dx_{2,3}}$.

\section{The performance of the Pseudo Spectral Analytical Time Domain (PSATD) solver}
\label{sec: psatd}
In this appendix, we check the associated EM fields of a free-streaming charged particle with the PSATD solver \cite{haber1973advances, birdsall1991plasma, vay2013domain}. The PSATD solver updates the Maxwell equations in the $\bm{k}$-space as \cite{vay2013domain}
\begin{align}
\bm{\tilde{E}}^{n+1} &= c \bm{\tilde{E}}^n + i s \hat{\bm{k}} \times \bm{\tilde{B}}^n - \frac{s}{k}\bm{\tilde{J}}^{n+1/2} + (1-c) 	\hat{\bm{k}}( \hat{\bm{k}} \cdot \bm{\tilde{E}}^n)  + \left( \frac{s}{k} - dt \right)\hat{\bm{k}}( \hat{\bm{k}} \cdot \bm{\tilde{J}}^{n+1/2})  \nonumber \\
\bm{\tilde{B}}^{n+1} &= c\bm{\tilde{B}}^n - i s \hat{\bm{k}} \times \bm{\tilde{E}}^n + i \frac{1-c}{k}\hat{\bm{k}} \times \bm{\tilde{J}}^{n+1/2}
\end{align}
where $c=\mathrm{cos}(k dt), s=\mathrm{sin} (kdt)$. Apply the Fourier transform in the time domain, the magnetic fields are
\begin{align}
\bm{\tilde{B}} = \frac{1}{w - c}\left(  - i s \hat{\bm{k}} \times \bm{\tilde{E}} + i \frac{1-c}{k}\hat{\bm{k}} \times \bm{\tilde{J}}\right)
\end{align}
where $w=\mathrm{exp}(-i\omega dt)$. Substitute it to the equation of the $\bm{\tilde{E}}$ fields, we can get
\begin{align}
\frac{w^2 - 2wc  +1 } { w- c } \bm{\tilde{E}} &- \frac{ ( 1+ w )(1-c)}{ w - c } \hat{\bm{k}}( \hat{\bm{k}} \cdot \bm{\tilde{E}}) = \left( \frac{s}{k} \frac{ w - 1}{ w - c} -dt \right) \hat{\bm{k}}( \hat{\bm{k}} \cdot \bm{\tilde{J}}) -  \frac{s}{k} \frac{w- 1}{w - c} \bm{\tilde{J}}
\end{align}
Combined with the Gauss's law, the fields can be solved as
\begin{align}
\tilde{E}_1 &= \left[   -\frac{i k_1}{k^2} + \frac{i (k_2^2+k_3^2)}{k_1 k^2}\frac{\mathrm{sin}(kdt)}{kdt} \frac{ 1-\mathrm{cos}(\omega dt) } { \mathrm{cos} (\omega dt) -\mathrm{cos} (k dt) }  \right] \tilde{\rho} \nonumber \\
\tilde{E}_2 &= -i\left[ 1 + \frac{\mathrm{sin}(kdt)}{kdt} \frac{1-\mathrm{cos}(\omega dt)}{\mathrm{cos}(\omega dt) - \mathrm{cos}(k dt) }\right] \frac{k_2}{k^2} \tilde{\rho} \nonumber \\
\tilde{E}_3 &=-i\left[ 1 + \frac{\mathrm{sin}(kdt)}{kdt} \frac{1-\mathrm{cos}(\omega dt)}{\mathrm{cos}(\omega dt) - \mathrm{cos}(k dt) }\right] \frac{k_3}{k^2} \tilde{\rho}  \nonumber \\
\tilde{B}_1 & = 0 \nonumber \\
\tilde{B}_2 & = i \frac{k_3}{k_1 k^2} \frac{\mathrm{sin}(\omega dt)}{dt} \frac{1-\mathrm{cos}(kdt)}{\mathrm{cos}(\omega dt) - \mathrm{cos} (k dt)} \tilde{\rho}\nonumber \\
\tilde{B}_3 &=  - i \frac{k_2}{k_1 k^2} \frac{\mathrm{sin}(\omega dt)}{dt} \frac{1-\mathrm{cos}(kdt)}{\mathrm{cos}(\omega dt) - \mathrm{cos} (k dt)} \tilde{\rho}
\end{align}
where the particles are assumed to drift along the $x_1$ direction. It can be shown the above expressions reduce to the physical ones when $dt \rightarrow 0$. 

Apply the inverse Fourier transformations and follow the same procedures as before, the $E_1$ field on the grids is
\begin{align}
{E}^n_{1, i_1 , i_2, i_3 }  &=   -\frac{1}{(2\pi)^3}  \int_{-\bm{k_g}/2}^{\bm{k_g}/2} \mathrm{d}\bm{k}  \sum_{\bm{\nu}}\left[   -\frac{i k_1}{k^2} + \frac{i (k_2^2+k_3^2)}{k_1 k^2}\frac{\mathrm{sin}(kdt)}{kdt} \frac{ 1-\mathrm{cos}(\beta k_1' dt) } { \mathrm{cos} (\beta k_1' dt) -\mathrm{cos} (k dt) }  \right]  S(\bm{k'})   \tilde{\rho}^0 (\bm{k}') \nonumber \\
 &\mathrm{exp}\left[ i k_1( i_1 dx_1  - \beta n dt)+ i k_2 i_2 dx_2 + i k_3 i_3 dx_3 \right] \mathrm{exp}( - i\beta \nu_1 k_{g1} n dt) 
 \label{eq: B5}
\end{align}

\begin{figure}[htbp]
\begin{center}
\includegraphics[width=0.5\textwidth]{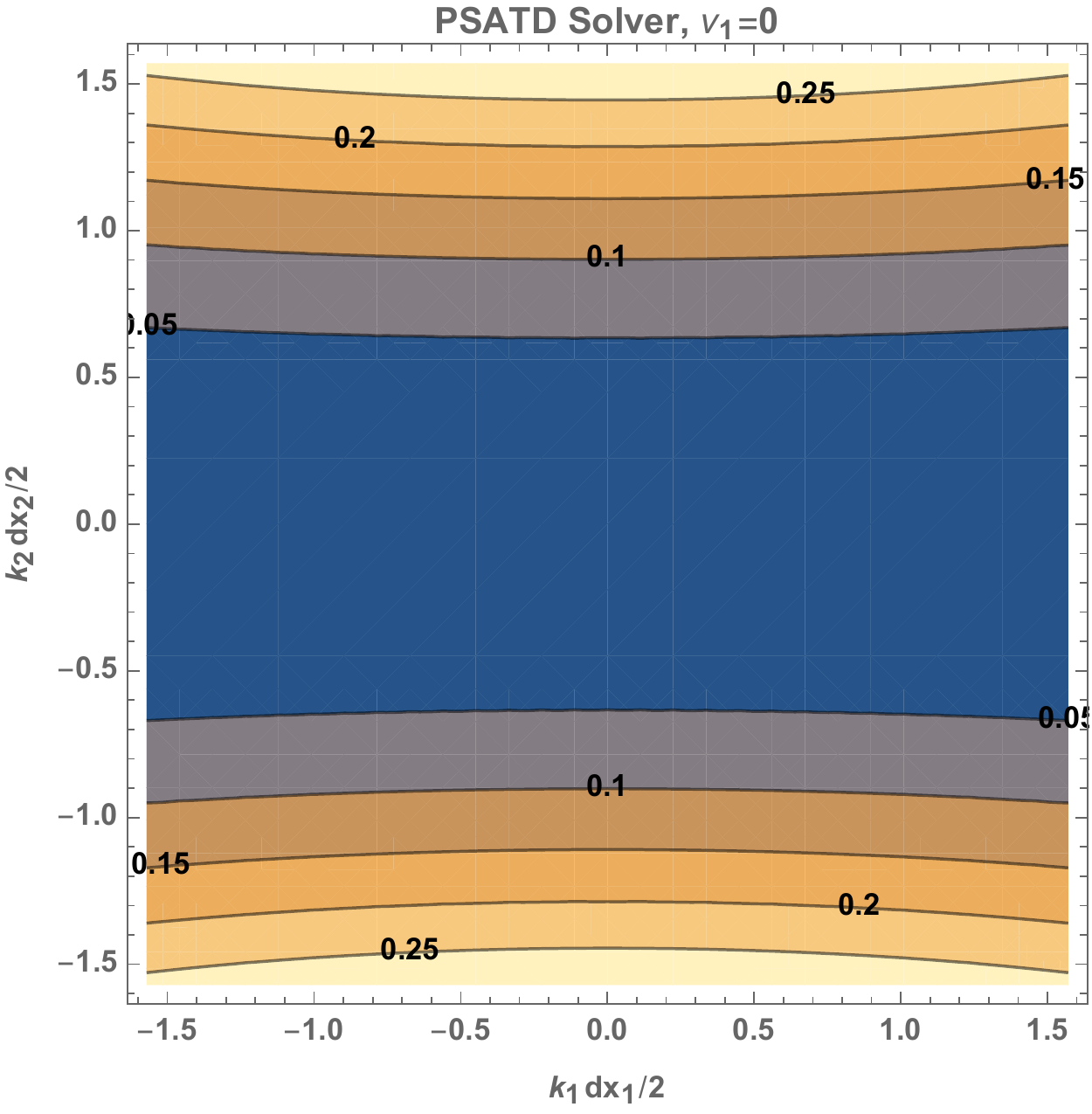}\includegraphics[width=0.5\textwidth]{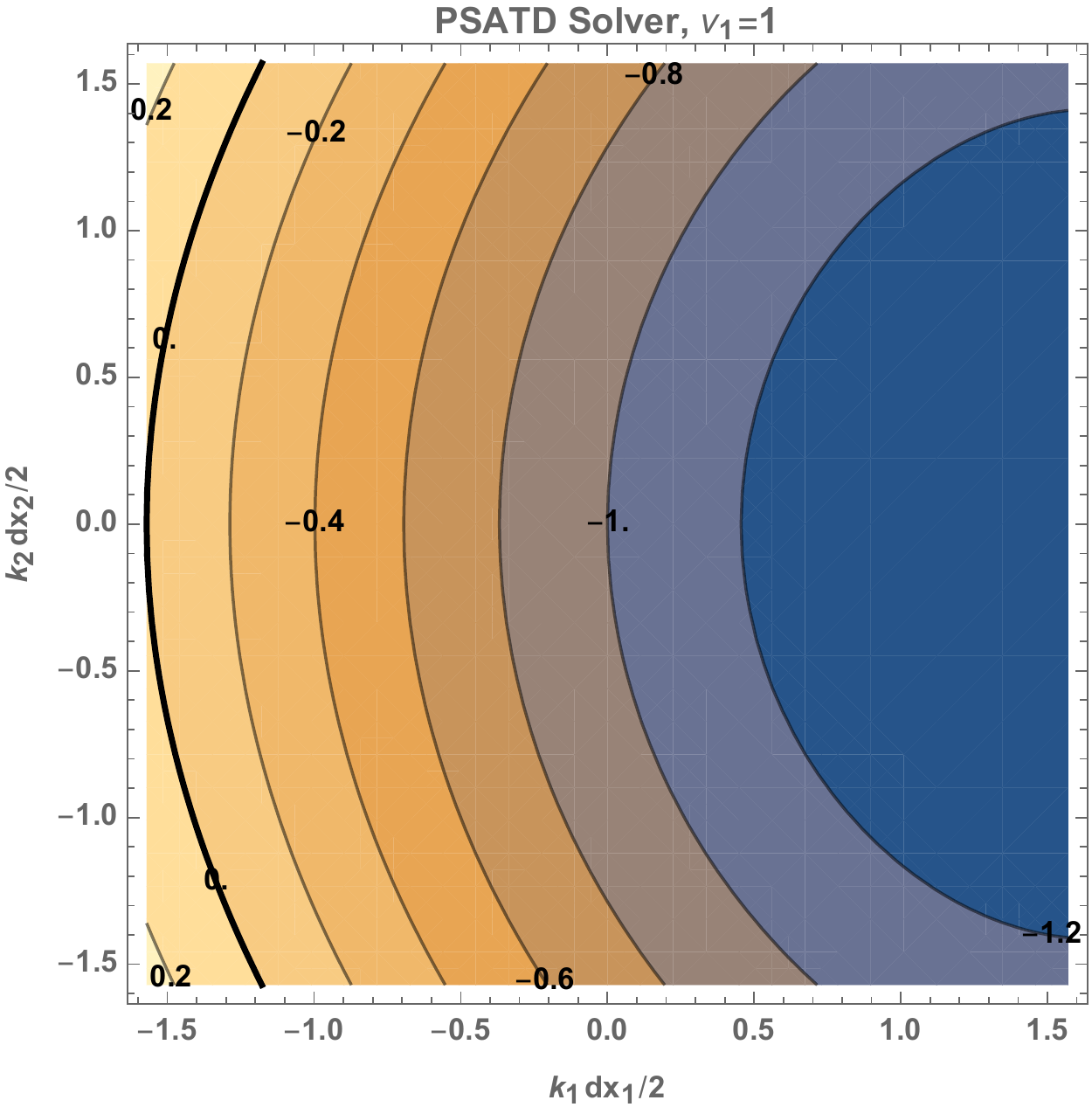}
\caption{The value of $\mathrm{cos} [ (k_1+\nu_1 k_{g1}) dt] -\mathrm{cos} (k dt)$ for the PSATD solver. The $E_1$ field from ZIPC simulations. Parameters: $dx_1=1, r\equiv \frac{dx_2}{dx_1}=1, \kappa\equiv \frac{dx_1}{dt}=4, q=1$.}
\label{fig: psatd}
\end{center}
\end{figure}

The value of the denominator of the integration function $\mathrm{cos} [ (k_1+\nu_1 k_{g1}) dt] -\mathrm{cos} (k dt)$ when $\nu_1=0$ and $\nu_1=1$ are shown in Fig. \ref{fig: psatd}. Its pattern is similar to the spectral solver. Thus we expect the numerical fields are also present in the PSATD solver. 

\section{Customizing stencil coefficients of arbitrary discrete operator $[k]_1$}
\label{sec: cus}

In this appendix, we will show how the method proposed in reference \cite{li2017controlling} is generalized to achieving constructing arbitrary discrete operator $[k_1]_\text{target}$. Arbitrary high order ($p$th order) finite difference operator with respect to $x_1$ has the form
\begin{align}
\partial^+_{x_1}f_{i_1,i_2}&=\frac{1}{\Delta x_1}\sum_{l=1}^M C_l^p(f_{i_1+l,i_2}-f_{i_1-l+1,i_2}), \\
\partial^-_{x_1}f_{i_1,i_2}&=\frac{1}{\Delta x_1}\sum_{l=1}^M C_l^p(f_{i_1+l-1,i_2}-f_{i_1-l,i_2}).
\end{align}
The corresponding operator in $k$-space becomes
\begin{equation}
[k_1]_p=\sum_{l=1}^M C_l^p\frac{\sin[(2l-1)k_1\Delta x_1/2]}{\Delta x_1/2}.
\end{equation}
For standard high order operator, the number of coefficients $M=p/2$. Here, in order to fit $[k_1]_p$ to an arbitrarily given $[k_1]_\text{target}$, we must have $M>p/2$ to give more freedom. For the simplicity of notations, we normalize $[k_1]_p$, $[k_1]_\text{target}$ and $k_1$ to $k_{g1}=2\pi/\Delta x_1$ herefrom. In the spirit of the least square approximation, such function should be minimized to obtain the stencil coefficients
\begin{equation}
\mathcal{F}=\int_0^{1/2}w(k_1)([k_1]_p-[k_1]_\text{target})^2 dk_1,
\end{equation}
where $w(k_1)$ is the weight function. In addition, the discrete operator is subjected to the constraint $\partial^\pm_{x_1}\rightarrow \partial_{x_1}+O(\Delta x_1^p)$, which can be guaranteed by the matrix equation $\mathcal{M}\vec{C}^p=\vec{e}_1$, where $\vec{C}^p\equiv(C_1^p,\cdots,C_M^p)^T$, $\vec{e}_1\equiv(1,0,\cdots,0)$ and the matrix element $\mathcal{M}_{ij}=(2j-1)^{2i-1}/(2i-1)!$ with $i=1,\cdots,p/2$ and $j=1,\cdots,M$. We use Lagrange multipliers to solve the constrained least-square minimization problem. The stencil coefficients are determined by minimizing the Lagrangian $\mathcal{L}\equiv\mathcal{F}+\vec{\lambda}^T(\mathcal{M}\vec{C}^p-\vec{e}_1)$, where $\vec{\lambda}$ is the multiplier. It can be shown straightforwardly the following minimization conditions
\begin{equation}
\frac{\partial\mathcal{L}}{\partial C_j^p}=0,\ j=1,\cdots,M\quad
\text{and}\quad
\frac{\partial\mathcal{L}}{\partial \lambda_i}=0,\ i=1,\cdots,p/2
\end{equation}
can be reformatted into a matrix equation
\begin{equation}
\begin{pmatrix}
\mathcal{A} & \mathcal{M}^T \\
\mathcal{M} & 0
\end{pmatrix}
\begin{pmatrix}
\vec{C}^p \\
\vec{\lambda}
\end{pmatrix}=\begin{pmatrix}
\vec{b} \\ \vec{e}_1
\end{pmatrix}
\end{equation}
where $A$ is an $M\times M$ matrix and $\vec{b}$ is an $M$-dimensional vector with the elements
\begin{align}
\mathcal{A}_{ij}&=\frac{2}{\pi^2}\int_0^{1/2}w(k_1)\sin[(2i-1)\pi k_1]\sin[(2j-1)\pi k_1] dk_1, \\
b_i&=\frac{2}{\pi}\int_0^{1/2}w(k_1)\sin[(2i-1)\pi k_1][k_1]_\text{target}dk_1.
\end{align}
With the matrix equation above, the stencil coefficients can be easily obtained. However, because of only finite number of coefficients, it is usually impossible to fit the target operator $[k_1]_\text{target}$ uniformly in the whole primary Brillouin zone $k_1\in[0,1/2]$. In this case, we need to set proper weight function $w(k_1)$ to relax the requirement. Usually, we can set a super-Gaussian weight function to require an accurate fit within the low and moderate $k_1$ region, and loose requirement for the high $k_1$ region
\begin{equation}
w(k_1)=\exp\left[-\ln2\left(\frac{2k_1}{w_{k1}}\right)^n\right]
\end{equation}
where $w_{k1}$ is the full width at half maximum of the weight function.

\section{Integrations in the complex plane}
\label{sec:integrations}
\begin{figure}[htbp]
\centering
\includegraphics[width=\textwidth]{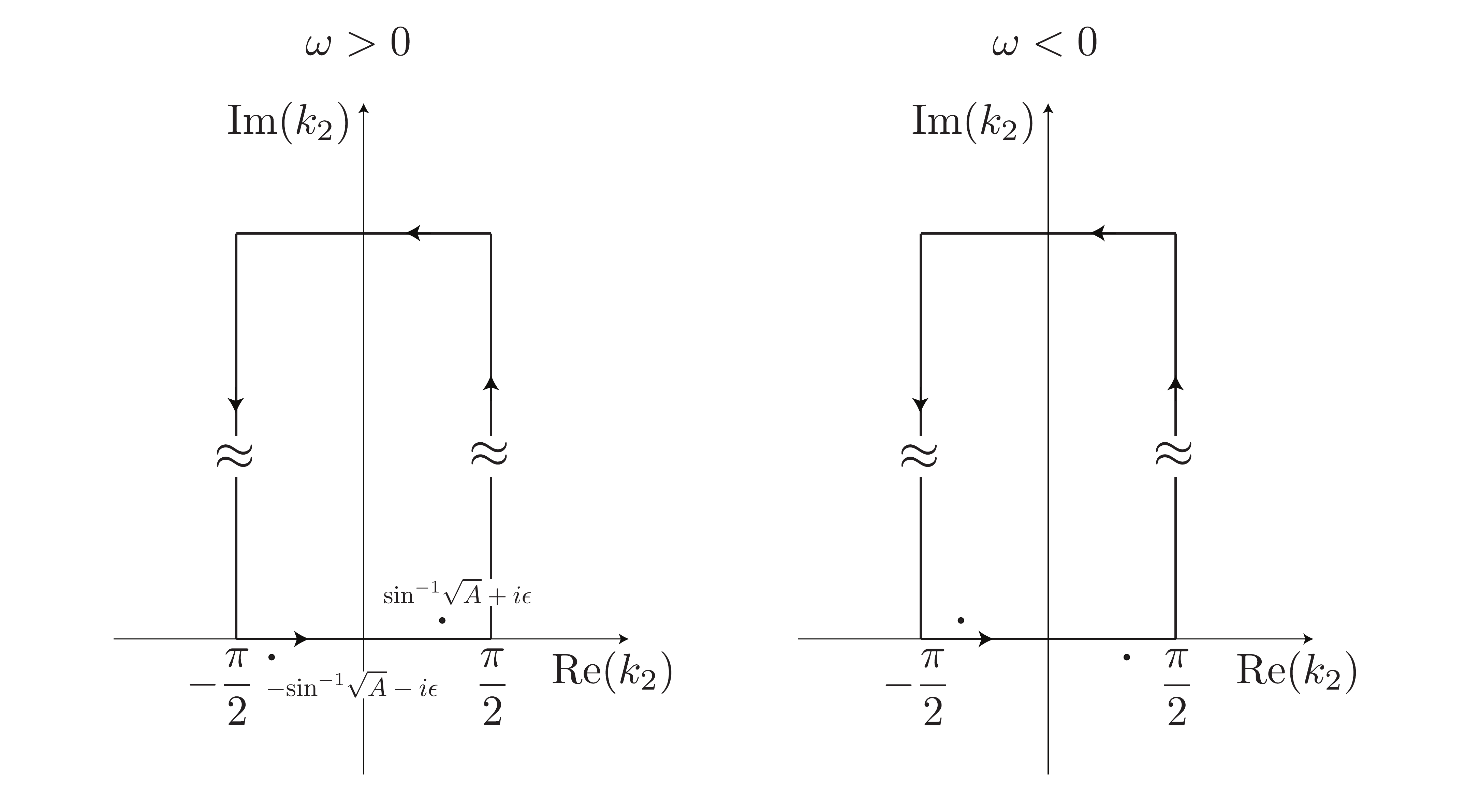}
\caption{The integration path and the locations of the poles when $k_1>0$ and $k_1<0$.}
\label{fig:path}
\end{figure}

In this appendix, we show how to do the integrations in the complex $k_2$ plane that are required to obtain the axial electric field for the different solvers. We first begin with Eq. 16. The integrand can be simplified as $\frac{A}{A-\mathrm{sin}^2 k_2}$, where $A>0$. To evaluate the integral we  use  the  closed path shown in Fig. \ref{fig:path} together with the residue theorem. 
\begin{align}
\ointop_C \mathrm{d} k_2 \frac{A}{A - \mathrm{sin}^2k_2} &= \left( \int_{bottom}  +  \int_{C_{right}} + \int_ {C_{top}} + \int_ {C_{left}} \right) \mathrm{d} k_2 \frac{A}{A - \mathrm{sin}^2k_2} \nonumber \\
& = \left( \int_{bottom}+   \int_ {C_{top}}  \right)  \mathrm{d} k_2 \frac{A}{A - \mathrm{sin}^2k_2} 
\end{align}
where it is straightforward to show the integrations along path $C_{left}$ and $C_{right}$ cancel each other, i.e., the integrand is equal for $k_{2,R}=k_g/2, k_{2,I}$ and $k_{2,R}=-k_g/2, k_{2,I}$. The integration along the line on the top is 
\begin{align}
 | \int_{C_{top}} \mathrm{d} k_2  \frac{A}{A-\mathrm{sin}^2k_2} | 	&= | \int_{-\pi/2} ^{\pi/2} \mathrm{d} k_{2,R}  \frac{A} { A +  \frac{ \mathrm{exp}(2 k_{2,I} )  \mathrm{exp}( -2 i  k_{2,R} ) - 2 + \mathrm{exp}(-2 k_{2,I} )  \mathrm{exp}( 2 i k_{2,R} )  }{4}} | \nonumber \\
&\leq \int_{-\pi/2} ^{\pi/2} \mathrm{d} k_{2,R}  \frac{A} { | A +  \frac{ \mathrm{exp}(2 k_{2,I} )  \mathrm{exp}( -2 i  k_{2,R} ) - 2 + \mathrm{exp}(-2 k_{2,I} )  \mathrm{exp}( 2 i k_{2,R} )   }{4} |} \nonumber \\
&\leq \int_{-\pi/2} ^{\pi/2} \mathrm{d} k_{2,R}   \frac{A} { | \frac{ \mathrm{exp}(2 k_{2,I} )  \mathrm{exp}(-2 i k_{2,R}) + \mathrm{exp} (-2 k_{2,I})  \mathrm{exp}( 2 i k_{2,R} )}{4} |  - |A -\frac{1}{2} |} 
\end{align} 
We can see this contribution is zero because along this path the integrand vanishes for all $k_{2,R}$ because $k_{2,I} \to +\infty$.

From causality, when $\omega >0$ which is equivalent to $k_1>0$ here,  there are two poles close to the $\mathrm{Re}(k_2)$ axis: one is at $k_2 = \mathrm{sin}^{-1}\sqrt{A} + i \epsilon$ and the other is at $k_2 = - \mathrm{sin}^{-1}\sqrt{A} - i \epsilon$; when $\omega<0$ ($k_1 <0$), there are two poles close to the $\mathrm{Re}(k_2)$ axis: one is at $k_2 = \mathrm{sin}^{-1}\sqrt{A} - i \epsilon$ and the other is at $k_2 = -\mathrm{sin}^{-1}\sqrt{A} + i \epsilon$, where $\epsilon \to 0$ from the positive side. 

We calculate the integration when $k_1 >0 $ first. Using the residue theorem, we get 
\begin{align}
\ointop_C \mathrm{d} k_2 \frac{A}{A - \mathrm{sin}^2k_2}  &= 2\pi i  \mathrm{Res} \left( \frac{A}{A - \mathrm{sin}^2k_2} , \mathrm{sin}^{-1}\sqrt{A} \right) \nonumber \\
& = 2\pi i \frac{A}{-2 \sqrt{A} \sqrt{1-A}} \nonumber \\
&= - i \pi \sqrt{\frac{A}{1-A^2}}
\end{align}
thus we know
\begin{align}
\int_{-\pi/2}^{\pi/2} \mathrm{d} k_2 \frac{A}{A - \mathrm{sin}^2k_2} = - i \pi \sqrt{\frac{A}{1-A^2}}
\end{align}

Similarly, the integration when $k_1<0$ can be obtained as
\begin{align}
\int_{-\pi/2}^{\pi/2} \mathrm{d} k_2 \frac{A}{A - \mathrm{sin}^2k_2} =  i \pi \sqrt{\frac{A}{1-A^2}}
\end{align}

The integration of Eq. 27 can be obtained similarly. 

Next we show how to do the integrations in Eq. 23. \begin{align}
\ointop_C \mathrm{d} k_2 \frac{A}{A - k_2^2} &= \left( \int_{bottom}  +  \int_{C_{right}} + \int_ {C_{top}} + \int_ {C_{left}} \right) \mathrm{d} k_2 \frac{A}{A - k_2^2} 
\end{align}

Again it can be shown the integration along the top line is zero where $k_{2I}\to\infty$,
\begin{align}
\mathrm{lim} _ {k_{2,I} \to +\infty} | \int_ {C_{top}} \mathrm{d} k_2 \frac{A}{A - k_2^2} |  & \leq \mathrm{lim} _ {k_{2,I} \to +\infty}  \int_{C_{top}} \mathrm{d} k_{2} \frac{A}{|A-k_2^2|} \nonumber \\
&\leq \mathrm{lim} _ {k_{2,I} \to +\infty}  \int_{C_{top}} \mathrm{d} k_{2} \frac{A}{|k_2^2| -A} = 0
\end{align}

The integrations along the two side lines do not cancel. Along the left side, the integration is
\begin{align}
\int_{C_{left}} \mathrm{d} k_2 \frac{A}{A - k_2^2} &= - \int_0^{+\infty} \mathrm{d} k_{2,I} i \frac{A}{A- (-\pi/2 + i k_{2,I} )^2}	\nonumber \\
&= -i \sqrt{A} \mathrm{tan}^{-1} \left( \frac{2 k_{2,I} + \pi i}{2\sqrt{A}}\right) \bigg | _{0}^{+\infty}	\nonumber \\
&= -i \sqrt{A} \left( \frac{\pi}{2} - i \mathrm{tanh}^{-1}\frac{\pi }{2\sqrt{A}}   \right)
\end{align}
The integration along the right line is
\begin{align}
\int_{C_{right}} \mathrm{d} k_2 \frac{A}{A - k_2^2} &=  \int_0^{+\infty} \mathrm{d} k_{2,I} i \frac{A}{A- (\pi/2 + i k_{2,I} )^2}	\nonumber \\
&= i \sqrt{A} \mathrm{tan}^{-1} \left( \frac{2 k_{2,I} - \pi i}{2\sqrt{A}}\right) \bigg | _{0}^{+\infty}	\nonumber \\
&= i \sqrt{A} \left( \frac{\pi}{2} + i\mathrm{tanh}^{-1}\frac{\pi }{2\sqrt{A}}   \right)
\end{align}
thus
\begin{align}
\left( \int_{C_{left}} + \int_{C_{right}}  \right) \mathrm{d} k_2 \frac{A}{A - k_2^2}  = -2 \sqrt{A} \mathrm{tanh}^{-1}\frac{\pi }{2\sqrt{A}} 
\end{align}

Using the residue theorem,
\begin{align}
\ointop_C \mathrm{d} k_2 \frac{A}{A - k_2^2} &= 2\pi i \mathrm{Res} \left( \frac{A}{A-k_2^2} , \sqrt{A} \right)	\nonumber \\
&= 2\pi i \frac{A}{-2 \sqrt{A}} = - \pi i \sqrt{A}
\end{align} 
Therefore, 
\begin{align}
\int_{-\pi/2} ^{\pi/2} \mathrm{d} k_2 \frac{A}{A-k_2^2} = -i\pi \sqrt{A} + 2 \sqrt{A} \mathrm{tanh}^{-1} \frac{\pi}{2\sqrt{A}}
\end{align}

\end{appendix}

\bibliographystyle{elsarticle-num}
\bibliography{refs_xinlu}

\begin{thebibliography}{10}
\bibitem{5passfilter}
It did a compensated 5 pass binomial smoothing, i.e. apply a $(1,2,1)$ stencil
  4 times, followed by a $(-5,14,-5)$ stencil.
\end{thebibliography}

\end{document}